\documentclass[12pt,a4paper]{article}
\usepackage[utf8x]{inputenc}
\usepackage{graphicx}
\usepackage{amsmath} 
\usepackage{amssymb} 

\usepackage[left=2cm,right=2cm,top=2cm,bottom=2cm]{geometry}
\usepackage[bookmarks=true,colorlinks=true,citecolor=blue,linkcolor=blue,
            urlcolor=magenta]{hyperref}
\usepackage[numbers,sort&compress]{natbib}

\title{Ultraslow Wave Nuclear Burning of Uranium-Plutonium Fissile Medium
       on Epithermal Neutrons}

\author{V.D.~Rusov$^1$\footnote{Corresponding author e-mail: siiis@te.net.ua}, 
        V.A.~Tarasov$^1$, M.V.~Eingorn$^2$, S.A.~Chernezhenko$^1$,\\ 
	A.A.~Kakaev$^1$, V.M.~Vashchenko$^3$, M.E.~Beglaryan$^1$}

\begin{document}

\maketitle

\noindent
$^1$Department of Theoretical and Experimental Nuclear Physics, 
Odessa National Polytechnic University, 
Shevchenko av. 1, Odessa 65044, Ukraine

\noindent
$^2$CREST and NASA Research Centers, 
North Carolina Central University, 
Fayetteville st. 1801, Durham, North Carolina 27707, U.S.A.

\noindent
$^3$State Ecological Academy for Postgraduate Education and Management 
Kyiv, Ukraine

\begin{abstract}
For a fissile medium, originally consisting of uranium-238, the investigation 
of fulfillment of the wave burning criterion in a wide range of neutron energies
is conducted for the first time, and a possibility of wave nuclear burning not 
only in the region of fast neutrons, but also for cold, epithermal and resonance
ones is discovered for the first time.

For the first time the results of the investigation of the Feoktistov criterion 
fulfillment for a fissile medium, originally consisting of uranium-238 dioxide 
with enrichments 4.38\%, 2.00\%, 1.00\%, 0.71\% and 0.50\% with respect to 
uranium-235, in the region of neutron energies 0.015$\div$10.00~eV are 
presented. These results indicate a possibility of ultraslow wave 
neutron-nuclear burning mode realization in the uranium-plutonium media, 
originally (before the wave initiation by external neutron source) having 
enrichments with respect to uranium-235, corresponding to the subcritical state,
in the regions of cold, thermal, epithermal and resonance neutrons.

In order to validate the conclusions, based on the slow wave neutron-nuclear 
burning criterion fulfillment depending on the neutron energy, the numerical 
modeling of ultraslow wave neutron-nuclear burning of a natural uranium in the 
epithermal region of neutron energies (0.1$\div$7.0~eV) was conducted for the 
first time. The presented simulated results indicate the realization of the 
ultraslow wave neutron-nuclear burning of the natural uranium for the 
epithermal neutrons.
\end{abstract}

\section*{Introduction}

Nowadays the development of the theory of the wave reactors with internal safety
(reactors of Feoktistov type)~\cite{ref01,ref02,ref03,ref04} as well as of the 
natural georeactor~\cite{ref03,ref04,ref05} is topical.

In~\cite{ref01} by an example of originally engineering uranium, irradiated by 
an external neutron source, for the arising uranium-plutonium fissile medium 
(the fertile nuclide $^{238}_{92}U$ and the fissile nuclide $^{239}_{94}Pu$) 
L.P.~Feoktistov proposed a criterion (a condition), the fulfillment of which in 
the neutron multiplication medium leads to a stationary wave of slow nuclear 
burning formation. The Feoktistov criterion consists in the condition that the 
equilibrium concentration of a fissile nuclide (for the considered neutron 
multiplication medium) must exceed its critical concentration. For the 
uranium-plutonium chain of nuclear reactions considered in~\cite{ref01}, by 
\textit{equilibrium} and \textit{critical} concentrations of the active 
component $^{239}_{94}Pu$ we mean the concentrations, for which the rate of the 
active component nuclei creation is equal to the rate of their disappearance 
during nuclear reactions, and the rate of neutron birth is equal to the rate of 
their absorption, respectively. For the particular case considered 
in~\cite{ref01}, this criterion reads: $N_{equil}^{^{239}_{94}Pu} > N_{crit}^{^{239}_{94}Pu}$. Since the computer modeling of the neutron multiplication medium 
kinetics is a very complicated problem, requiring large computational burden, 
the check of the Feoktistov criterion becomes the only way of the preliminary 
search for nuclide composition of the neutron multiplication medium and the 
external parameters, for which the realization of a slow nuclear burning wave is
possible. However, conducting of such a search is even more complicated by the 
fact that the equilibrium and critical concentrations of the fissile nuclide, 
present in the Feoktistov criterion, are functions of the energy spectrum of 
neutrons, which changes with the nuclide composition and the external parameters
such as the temperature, the pressure and geometry of a fissile medium.

\section{Akhiezer wave of nuclear burning for the thermal neutrons}

Although in~\cite{ref02} the other problem was solved and investigated, it is 
of undoubtful interest for our subject and makes a certain contribution into 
theoretical generalization of the possible neutron multiplicative burning 
processes and substantiation of the possibility of the wave burning realization 
in the thermal region of neutron energies in principle.

A chain nuclear reaction on thermal neutrons in the large neutron-multiplication
system in the form of a cylinder or a parallelepiped of big length (one of the 
geometric parameters must exceed two others considerably) was considered in the 
one-group diffusion-age approximation in~\cite{ref02}. The chain nuclear 
reaction was initiated by the external neutron source, which was set by its 
neutron density flux on the cylinder butt (the problem II in~\cite{ref02}).

At the same time in~\cite{ref02} the thermal neutron density $n (\vec{r},t)$ 
satisfies the following diffusion integro-differential equation, which can be 
derived from the effective one-group equation in the diffusion-age 
approximation~\cite{ref06,ref07,ref08}, taking into account that the so-called 
age of thermal neutrons $\tau$ is proportional to the mean value of the squared 
neutron shift during the deceleration process from the point of fast neutron 
birth to the point of thermal neutron birth (see e.g.~\cite{ref07,ref08,ref09}):

\begin{equation}
\frac{\partial n(\vec{r},t)}{\partial t} = D \Delta n (\vec{r},t) - 
\frac{1}{\tau_c} n (\vec{r},t) + \frac{k_\infty}{\pi ^{3/2}r_0 ^3 \tau_c} 
\int \limits_V n(\vec{r}',t) exp \left( - \frac{\left \vert \vec{r} - \vec{r}' 
\right \vert^2}{r_0 ^2} \right) dV'
\label{eq01}
\end{equation}

\noindent
where $D$ is the diffusion coefficient for thermal neutrons, $\tau_c$ is their 
lifetime with respect to capture, $k_\infty$ is the neutron infinite 
multiplication factor, $r_0$ is the mean neutron moderation length.

The form of the integrand in~(\ref{eq01}) supposes that all dimensions of a 
neutron-multiplication system are considerably larger than the neutron 
moderation length $r_0$. Let us note that although integration in~(\ref{eq01}) 
is performed over the volume occupied by the neutron-multiplication medium, in 
fact, because of the smallness of the mean neutron moderation length $r_0$ in 
comparison with typical dimensions of the system, it covers only a small 
neighbourhood of the point $\vec{r}'$ with the diameter of the order of $r_0$. 
This allows to extend the limits of integration in~(\ref{eq01}), containing the 
Gaussian function, to infinity with exponential accuracy.

As of the neutron multiplication factor, it was set in~\cite{ref02} in the form 
of the known four factor formula for thermal reactors (see 
e.g.~\cite{ref07,ref08,ref09}) and neglecting the fast multiplication factor 
$\varepsilon$, which is close to unity for thermal reactors ($\varepsilon \cong 
1.00 \div 1.03$):

\begin{equation}
k_\infty = \nu \varepsilon \theta _f \phi \approx \nu \theta _f \phi
\label{eq02}
\end{equation}

\noindent
where $\nu$ is the average number of neutrons, generated under capture of one 
neutron by the fuel, $\theta _f$ is a probability of thermal neutron absorption 
by an uranium nucleus, $\phi$ is the probability function for the resonance 
neutron absorption by uranium-238 nuclei.

However, in order to simplify the problem, only the kinetic equation for 
neutrons was considered, and the kinetic equations for densities of nuclides in 
the multiplication fissile medium were not taken into account. Otherwise one had
to solve the system of kinetic equations with 
nonlinear feedbacks
(an example of such kinetic system of 20 equations is presented below in 
section~\ref{sec05}, as well as e.g. in~\cite{ref01,ref03,ref04,ref05,ref10}).

Really, the neutron multiplication factor in~\cite{ref02} was set at the level 
of the expression~(\ref{eq02}), although, as is known (see 
e.g.~\cite{ref07,ref08,ref09}), each of the factors present in~(\ref{eq02}) 
depends on the composition of the fuel fissile medium and its construction, for 
example, $\nu$ is set by the following expression:

\begin{equation}
\nu = \frac{\sum \limits _i \nu_i \sigma_f ^i N_i}{\sum \limits _i \sigma_a ^i N_i}
\label{eq03}
\end{equation}

\noindent
where $\nu_i$ is the average number of neutrons, generated by the fission of one
nucleus of the $i^{th}$ fissile nuclide, $\sigma_f ^i$ is the fission 
cross-section of the $i^{th}$ nuclide, $\sigma _a ^i$ is the neutron absorption 
cross-section of the $i^{th}$ nuclide, $N_i$ is the density of the $i^{th}$ 
nuclide nuclei.

Thus, in~\cite{ref02} the change of the fissile medium composition was not taken
into account, and the process of neutron multiplication was described at the 
level of the neutron multiplication factor (the expressions (\ref{eq01}) and 
(\ref{eq02})), which is known to depend on the nuclide composition of the 
fissile medium and cross-sections of neutron-nuclear reactions (the expression 
(\ref{eq03})). The cross-sections, in their turn, depend on the neutron energy 
and can change their values considerably (see e.g. figures~\ref{fig:01} 
and~\ref{fig:02}). Therefore, under such a simplified description, the 
dependences on the fissile medium composition and the neutron energy were 
at the back of the multiplication factor. Naturally, this imposed 
corresponding restrictions on the solutions obtained and conclusions made in~\cite{ref02}.

The solution of the kinetic equation for the density of thermal neutrons~(\ref{eq01}) was obtained in~\cite{ref02} for a semi-infinite parallelepiped 
($0 \leqslant z \leqslant +\infty$) of square section 
($0 \leqslant x,y \leqslant a$), on the butt of which ($z=0$) either the flux 
$j_0 (x,y,t)$, or the neutron density $n_0 (x,y,t)$ was set, corresponding to 
presence of the external neutron source at the butt (the problem II 
in~\cite{ref02}). The investigation of its behavior for asymptotics with respect
to time and parallelepiped length was also carried out for two particular cases 
of the above-critical and subcritical states of the neutron multiplication 
fissile medium.

Let us note that the choice of geometry of a fissile medium in the form of a 
cylinder or a parallelepiped of large length was made in~\cite{ref02}, because 
in this case the above-critical or subcritical states practically do not depend 
on the length, i.e. one can neglect the influence of butts, and the ratio of the
surface area to the volume remains practically constant with the length change.

Following~\cite{ref02}, if the system is surrounded by vacuum, and its 
dimensions are large in comparison with the neutrons free path, the boundary 
condition at the external lateral surface ($x = 0 \div a$; $y = 0\div a$) 
consists in the equality of the neutron density to zero. Thus, we look for the 
solutions of~(\ref{eq01}) in the following form:

\begin{equation}
n(r,t) = n(z,t) \sin (\pi x / a) \sin (\pi y / a)
\label{eq04}
\end{equation}

Let us note that, of course, the zero neutron flux density should be set on the 
extrapolated lateral surface of the system, but this inaccuracy in~\cite{ref02} 
may be disregarded, since it can be easily corrected.

Substituting~(\ref{eq04}) into the equation~(\ref{eq01}) we get the following 
equation for the function $n(z,t)$:

\begin{equation}
\frac{\partial (z,t)}{\partial t} = D \frac{\partial ^2 n(z,t)}{\partial z^2} - 2 \frac{\pi^2 D}{a^2} n(z,t) - \frac{1}{\tau_c} n(z,t) + \frac{k_\infty}{\pi ^{1/2}r_0 \tau_c} e^{-\frac{\pi^2 r_0 ^2}{2 a^2}} \int \limits _0 ^\infty n(z',t) exp \left( - \frac{(z - z')^2}{r_0 ^2} \right) dz'
\label{eq05}
\end{equation}

Since the diffusion flux reads $j_z = -D \frac{\partial n(z,t)}{\partial z}$, 
the boundary and initial conditions for the problem II in~\cite{ref02} have the 
form:

\begin{equation}
n(z,t) \vert _{t=0} = 0, ~~ n(z,t) \vert _{z=0} = n_0 (t) ~~ or ~~ 
\left. \frac{\partial n(z,t)}{\partial z} \right \vert _{z=0} = -\frac{j_0 (t)}{D}
\label{eq06}
\end{equation}

At the same time the boundary conditions on the butt of the cylinder were as 
follows:

\begin{equation}
j_0 (x,y,t) = j_0 (t) \sin (\pi x/a) \sin (\pi y/a); ~~
n_0 (x,y,t) = n_0 (t) \sin (\pi x/a) \sin (\pi y/a)
\label{eq07}
\end{equation}

The solution of the equation~(\ref{eq05}) with the initial and boundary 
conditions~(\ref{eq06}) and~(\ref{eq07}) (the problem II in~\cite{ref02}) with 
the help of the direct and inverse Fourier transforms was obtained 
in~\cite{ref02} in the following form:

\begin{equation}
n(z,t) = \frac{1}{\sqrt{\pi D^*}} \int \limits _0 ^t j_0 (t - \mu) \frac{\exp \left( \frac{A^*}{\tau_c} \mu - \frac{z^2}{4D^* \mu} \right)}{\sqrt{\mu}} d \mu,
\label{eq08}
\end{equation}

\noindent where

\begin{equation}
D^* = D + \frac{1}{4} k_\infty \frac{r_0 ^2}{\tau_c}
\label{eq09}
\end{equation}

\noindent is called the effective diffusion coefficient~\cite{ref06},

\begin{equation}
A^* = k_\infty - 1 - \frac{2 \pi^2}{a^2} D^* \tau_c
\label{eq10}
\end{equation}

\noindent is the effective neutron multiplication factor.

In~\cite{ref02} the obtained solution~(\ref{eq08}) was studied assuming that the
neutron flux is constant at the boundary of the right-angled cylinder for $z=0$,
i.e. it does not depend on time. In this case the solution~(\ref{eq08}) is 
simplified and has the following form:

\begin{equation}
n(z,t) = \frac{j_0}{\sqrt{\pi D^*}} \int \limits _0 ^t \frac{\exp \left( \frac{A^*}{\tau_c} \mu - \frac{z^2}{4D^* \mu} \right)}{\sqrt{\mu}} d \mu = \frac{j_0}{\sqrt{\pi D^*}} J
\label{eq11}
\end{equation}

\noindent
where $J$ represents a designation for the integral in~(\ref{eq11}).

Authors of~\cite{ref02} were first of all interested in the above-critical mode
$A^* > 0$, and it was shown that in this case one can speak about some velocity 
of slow nuclear burning. In order to find this velocity, asymptotics of the 
expression~(\ref{eq11}) for were considered for $z = v t$, $v = const$, 
$z,t \to \infty$, i.e. in observation points moving along the body axis with one
or another constant velocity $v$. The asymptotics of $n(z,t)$ are easily 
reproduced, if one transforms the integral $J$ in~(\ref{eq11}) via the 
substitution of variables to the form of the standard Laplace integral 
$\int \limits _a ^b \varphi (u) \exp ({\lambda u}) du$, where $\lambda$ is a 
large positive parameter, and $u$ is a real number~\cite{ref02}.

As shown in~\cite{ref02}, if one introduces the quantities

\begin{equation}
v_0 = 2 \sqrt{\frac{A^* D^*}{\tau_c}}
\label{eq12}
\end{equation}

\noindent and

\begin{equation}
L_0 = 2 \sqrt{\frac{D^* \tau_c}{A^*}} = \frac{4D^*}{v_0}
\label{eq13}
\end{equation}

\noindent
having dimensionalities of velocity and length, then, if the quantity $2z / L_0$
in the exponential function in the integral $J$, transformed to the form of the 
standard Laplace integral, is a large parameter ($z/L_0 \gg 1$), integrating $J$
by parts, we find the main term of asymptotics of the integral $J$ and, 
accordingly, the asymptotics of the expression~(\ref{eq11}) for the neutron 
density $n(z,t)$ for $z = v t \to \infty$, $v = const$ in the following 
form:

\begin{align}
n(z,t) & \approx \frac{j_0}{2\sqrt{\pi D^*}} \frac{L_0}{z} \exp 
\left[ \frac{z}{L_0} \left( \frac{v_0}{v} - \frac{v}{v_0} \right) \right] 
\varphi \left( \frac{1}{2} \left( \frac{v_0}{v} - \frac{v}{v_0} \right) \right) = {} \nonumber \\
{} & = \frac{4 j_0}{\sqrt{\pi z}} \sqrt{D^*} \exp \left[ \frac{z}{L_0} \left( \frac{v_0}{v} - \frac{v}{v_0} \right) \right] \frac{v ^{1/2}}{v^2 + v_0 ^2}
\label{eq14}
\end{align}

As it is seen from~(\ref{eq14}), for all constant velocities $v < v_0$ the 
quantity $n(z,t)$ exponentially grows up with the increase of the distance $z$ 
from the cylinder butt (or with time, since $z = v t$). At the same time in 
the case $v > v_0$, on the contrary, the quantity $n(z,t)$ exponentially 
decreases with the increase of the distance $z$. If the velocity $v$ is equal 
to the velocity $v_0$, then an abrupt change of the asymptotics type of 
$n(z,t)$ occurs: instead of being exponential, it becomes power and decreases 
very slowly with distance:

\begin{equation}
n(z,t) \approx \frac{j_0}{\sqrt{\pi}v_0} \sqrt{\frac{L_0}{z}}
\label{eq15}
\end{equation}

Using the expression~(\ref{eq14}), one can find the velocity of the constant 
thermal neutron density propagation at large distances from the cylinder butt. 
Supposing $n(z,t) = \tilde{n}$, we seek for the solution for the velocity $z/t$ 
in the following form:

\begin{equation}
\frac{z}{t} = v_0 + \varepsilon (z), ~~ \frac{\varepsilon (z)}{v_0} \ll 1
\label{eq16}
\end{equation}

As a result, we get

\begin{equation}
\frac{\varepsilon (z)}{v_0} \approx \frac{1}{2} \frac{L_0}{z} \ln \left[ \frac{j_0}{\sqrt{\pi} \tilde{n} v_0} \sqrt{\frac{L_0}{z}} \right], ~~ \frac{z}{L_0} \gg 1
\label{eq17}
\end{equation}

Differentiating~(\ref{eq16}) with respect to time $t$, we find the following 
expression for the instantaneous velocity $dz / dt$:

\begin{equation}
\frac{dz}{dt} = \frac{v_0 + \varepsilon(z)}{1 - t \frac{d \varepsilon}{dz}} \approx v_0 \left( 1 - \frac{L_0}{4z} \right)
\label{eq18}
\end{equation}

Thus, the instantaneous velocity of the constant thermal neutron density 
propagation in the first order of $L_0 / z$ does not depend on the neutron 
density $\tilde{n}$ and at large distances from the cylinder butt tends 
asymptotically to $v_0$ (the velocity of slow nuclear burning in a fissile 
medium). Really, in the case of usual slow burning, when a certain temperature 
is reached as a result of the reaction, the slow burning velocity is 
proportional to $\sqrt{\lambda / \tau}$, where $\lambda$ is the thermal 
diffusivity, and $\tau$ is the characteristic time of the 
reaction~\cite{ref11,ref12}. In the considered case it is a question of 
achievement of some fixed neutron density at the given point, caused by 
multiplication as well as by neutron diffusion. According to the 
expression~(\ref{eq12}) for $v_0$, the lifetime of a neutron $\tau_c$ plays 
the role of a characteristic time of the reaction, while the geometric mean 
from the effective diffusion coefficient $D^*$~(\ref{eq09}) and the effective 
multiplication coefficient $A^*$~(\ref{eq10}) plays the role of the transport 
factor.

In~\cite{ref02} the asymptotics of the neutron density $n(z,t)$ were also 
considered in two important cases (of large distances and large time lapses), 
and the following asymptotic expressions were obtained:

\begin{itemize}
\item in the case of growing $z$ and fixed $t$, i.e. when $z \to \infty$, $t = const$ and $v / v_0 \gg 1 (v \neq const)$

\begin{equation}
n(z,t) \approx \frac{4j_0}{\sqrt{\pi}} \frac{\sqrt{D^* t^3}}{z^2} e^{A^* t/\tau_c} e^{-z^2 / (4D^* t)} , ~~ \frac{z^2}{4D^* t} \gg 1;
\label{eq19}
\end{equation}

\item in the case of growing $t$ and fixed $z$, i.e. when $t \to \infty$, $z = const$ and  $v / v_0 \ll 1$

\begin{equation}
n(z,t) \approx \frac{2j_0}{\sqrt{\pi}v_0} \frac{1}{\sqrt{A^* \frac{t}{\tau_c}}} e^{A^* t/\tau_c} , ~~ A^* \frac{1}{\tau_c} \gg 1 , ~~ v = \frac{z}{t} .
\label{eq20}
\end{equation}

\end{itemize}

The first of the considered cases corresponds to the ``instantaneous'' picture 
of neutron density distribution over the whole cylinder length, and the second 
one describes the density evolution at each given fixed point $z$. From the 
expressions for asymptotics~(\ref{eq19}) and~(\ref{eq20}) it is clear that at 
large distances the quantity $n(z,t)$ decreases exponentially with distance, and
for large lapses of time for the above-critical mode ($A^* > 0$) it grows up 
exponentially with time because of multiplication. It is also clear that for 
$A^* > 0$ the neutron density grows up with time unlimitedly, and the quantity 
$\frac{\tau_c}{A^*}$ is a typical time. This means that in this case the chain 
reaction of spontaneous fissions under impossibility of neutron and energy 
removal leads to a system explosion. As noted in~\cite{ref02}, it is necessary, 
however, to keep in mind that the exponential growth of $n(z,t)$ is related 
essentially to linearity of the used approximation, and if the nonlinear terms 
with respect to $n(z,t)$ are taken into account, the growth intensity decreases.
The authors of~\cite{ref02} supposed that the neutron density $n(z,t)$ is a 
finite function, and the velocity of propagation of the burning front is the 
same as in the linear approximation.

In the case of the subcritical mode $A^* < 0$, when the number of neutrons, 
being born during the nuclei fission, is not enough for maintaining the 
spontaneous chain nuclear reaction, the exponents in the 
expressions~(\ref{eq19}) and~(\ref{eq20}) are negative and the neutron density 
decreases with time as well as with the increase of the distance from the 
cylinder butt. In this case a characteristic length of neutron propagation may 
be considered. The quantity $\sqrt{D^* \tau_c / \vert A^* \vert}$ represents 
such length, according to~\cite{ref02}.
The most important conclusion for us, made in~\cite{ref02} as a result of the 
analysis of behaviour for asymptotics of the solution~(\ref{eq11}) for the 
neutron density $n(z,t)$ and asymptotics with respect to time and the cylinder 
length~(\ref{eq19}) and~(\ref{eq20}), consists in the fact that for the case of 
the above-critical state of the neutron-multiplication fissile medium, a 
possible existence of the wave burning, propagating with the constant velocity 
and practically unchangeable amplitude along the cylinder axis and representing 
superposition of processes of multiplication and diffusion of neutrons, is 
demonstrated, and in the case of the subcritical state of the 
neutron-multiplication fissile medium, the neutron process decays. It is also 
important, as it is shown in~\cite{ref02}, that the wave burning velocity is 
defined by the formula similar to usual slow burning~\cite{ref11,ref12}. The 
estimate of this wave burning velocity, presented in~\cite{ref02}, amounts to 
$\sim$100~sm/s. Let us also note that the analysis of the asymptotics behaviour 
with respect to time and the cylinder length showed that for the case of the 
above-critical state of the neutron-multiplication fissile medium, for the 
values of velocities of the chain process propagation bigger than the aforesaid 
wave process velocity, the chain process run-away, i.e. the explosion, begins, 
and for the values of velocities of the chain process propagation, smaller than 
the aforesaid wave process velocity, the chain process decays. In~\cite{ref02} 
conditions and criteria, under which the wave burning mode, that may be called 
an above-critical Akhiezer wave of slow nuclear burning, can be realized, are 
absolutely not investigated, and the only condition, designated in this paper, 
is the initial above-critical state of the fissile medium for the whole 
cylinder. Of course, it is also very interesting to investigate the features of 
this wave kinetics, its stability, the degree of burn-up for fuel medium 
nuclides, etc.

Thus in~\cite{ref02} the uranium-plutonium~\cite{ref01,ref02,ref03,ref04,ref05} 
and thorium-uranium (see e.g.~\cite{ref02,ref03,ref04,ref05,ref10}) chains of 
nuclear reactions, underlying slow wave burning (the slow Feoktistov wave), on 
which the whole conception of wave reactors of new generation is based and due 
to which these reactors have the property of internal safety and allow to use 
the originally unenriched or slightly enriched media of uranium-238 or 
thorium-232 as fuel, were not considered. For slow wave nuclear burning, being 
the subject of the given paper, it is essential that the multiplication fissile 
uranium-plutonium or thorium-uranium (or even thorium-uranium-plutonium) medium 
was initially in the subcritical state, and as it follows from the results 
of~\cite{ref02}, in this case the wave chain process is not discovered.

Let us also note that the velocity of the slow Feoktistov wave is several orders
of magnitude smaller than the velocity of the Akhiezer wave, and therefore the 
Feoktistov wave may be called a wave of ultraslow neutron-nuclear burning. 
Besides, the burning region in the Feoktistov wave passes temporarily into the 
above-critical state as a result of accumulation of the fissile nuclide 
(plutonium-239 or uranium-233 for the thorium-uranium cycle, or both 
simultaneously), and after that the total process of multiplication and 
diffusion of neutrons, apparently, becomes similar to the chain neutron process,
considered in~\cite{ref02}, however, here the differences are also noticeable, 
since in the case of the Feoktistov wave this supercriticality is local, and 
for the above-critical Akhiezer wave supercriticality is set initially for the 
whole cylinder of the fissile medium.

It is natural to note that Feoktistov and Akhiezer criteria of the wave burning 
realization, as it follows from the present paper, differ essentially, that 
also indicates that these wave processes are different.

In connection with the aforesaid it is interesting to conduct the following 
computer modelling experiment: in the cylindrical uranium (or uranium-plutonium)
neutron-multiplication medium, being in the above-critical state with respect to
uranium-235 (or plutonium-239), with the help of the permanent external neutron 
source to initiate the Akhiezer burning wave (the wave of uranium-235 burning), 
after run of which along the cylinder by the condition of non-hundred-percent 
burn-up of uranium-238 (we make this stipulation because the kinetics of the 
Akhiezer wave is not studied for the time being) after some time, typical for 
lighting of the Feoktistov wave, the external neutron source will initiate the 
Feoktistov wave, which will also propagate along the cylinder. Of course, 
conditions of such a computer experiment must be agreed with criteria of 
existence of these wave processes, and first of all with respect to neutron 
energies. In this connection let us note that from physical considerations it is
clear that the Akhiezer wave by the corresponding initial enrichment with 
respect to the fissile nuclide, ensuring the above-critical state of the fissile
medium, can be realized not only for thermal neutrons, as considered 
in~\cite{ref02}, but also for other neutron energies, e.g. for fast neutrons. 
This is also confirmed by the fact that the used in~\cite{ref02} equation for 
thermal neutrons in the one-group diffusion-age approximation can be also 
generalized for other neutron energies (see 
e.g.~\cite{ref08,ref13,ref14,ref15}).

A similar computer experiment may be conducted also for thorium-uranium or 
thorium-uranium-plutonium fissile media.)

\section{Fulfillment of Feoktistov's criterion for uranium-plutonium neutron 
         multiplication medium and the neutron energy of 0.1~eV$\div$1~MeV}

\label{sec02}

In~\cite{ref01} for the uranium-plutonium medium under a number of 
simplifications of the kinetic system of equations for the considered process 
(the one-dimensional medium and the fixed neutron energy (the one-group 
approximation) are considered, the neutron diffusion is not taken into account, 
the kinetic equation for plutonium-239 is written assuming that uranium-238 
turns directly into plutonium-239 with some typical time of the 
$\beta$-transition $\tau_\beta$, delayed neutrons and the fissile medium 
temperature are not taken into account) the following expressions for the 
equilibrium concentration $N_{eq}^{Pu}$ of the fissile nuclide $^{239}_{94}Pu$ 
and its critical concentration $N_{crit}^{Pu}$ are obtained:

\begin{equation}
N_{eq}^{Pu} (E_n) \approx \frac{\sigma _c ^8 (E_n)}{\sigma _c ^{Pu} (E_n) + \sigma _f ^{Pu} (E_n)} N^8 = \frac{\sigma _c ^8 (E_n)}{\sigma _a ^{Pu} (E_n)} N^8
\label{eq21}
\end{equation}

\begin{equation}
N_{crit} ^{Pu} (E_n) \approx \frac{\sum \limits_{i \neq Pu} \sigma _a ^i (E_n) N^i - \sum \limits_{i \neq Pu} \nu_i \sigma_f ^i (E_n) N^8}{(\nu_{Pu} - 1) \sigma_f ^{Pu} (E_n) - \sigma_c ^{Pu} (E_n)}
\label{eq22}
\end{equation}

\noindent
where $\sigma_c ^i$, $\sigma_f ^i$, $\sigma_a ^i$ are micro-cross-sections of 
the neutron radiative capture reactions, fission and neutron absorption, 
respectively, for the $i^{th}$ nuclide of the fissile medium; $\tau_\beta$ is a 
typical time for two $\beta$-decays, transforming $^{239} _{92}U$ (arising under
the radiation capture of neutrons by uranium $^{238}_{92}$) into 
$^{239}_{93}Np$, and $^{239}_{93}Np$ into $^{239}_{94}Pu$; $\nu_i$ and 
$\nu_{Pu}$ represent average numbers of neutrons being born as a result of 
fission of one nucleus of the one $i^{th}$ nuclide and $^{239}_{94}Pu$, 
respectively.

\begin{figure}[htb]
\centering
\includegraphics[width=16cm]{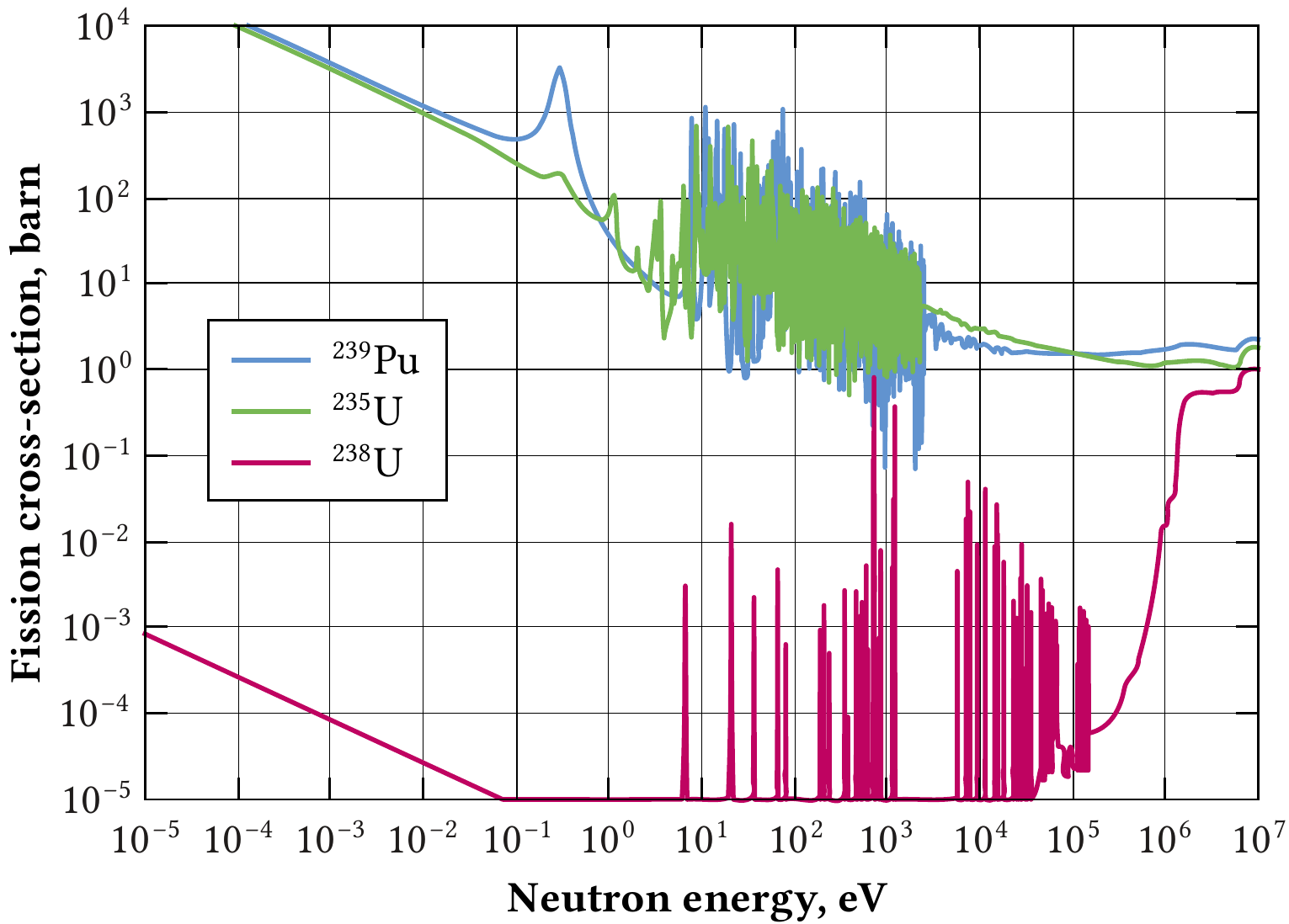}
\caption{The dependences of fission cross-sections on the energy of neutrons for
         $^{239}Pu$, $^{235}U$ and $^{238}U$, taken from the database 
	 ENDF/B-VII.0.}
\label{fig:01}
\end{figure}

\begin{figure}[htb]
\centering
\includegraphics[width=16cm]{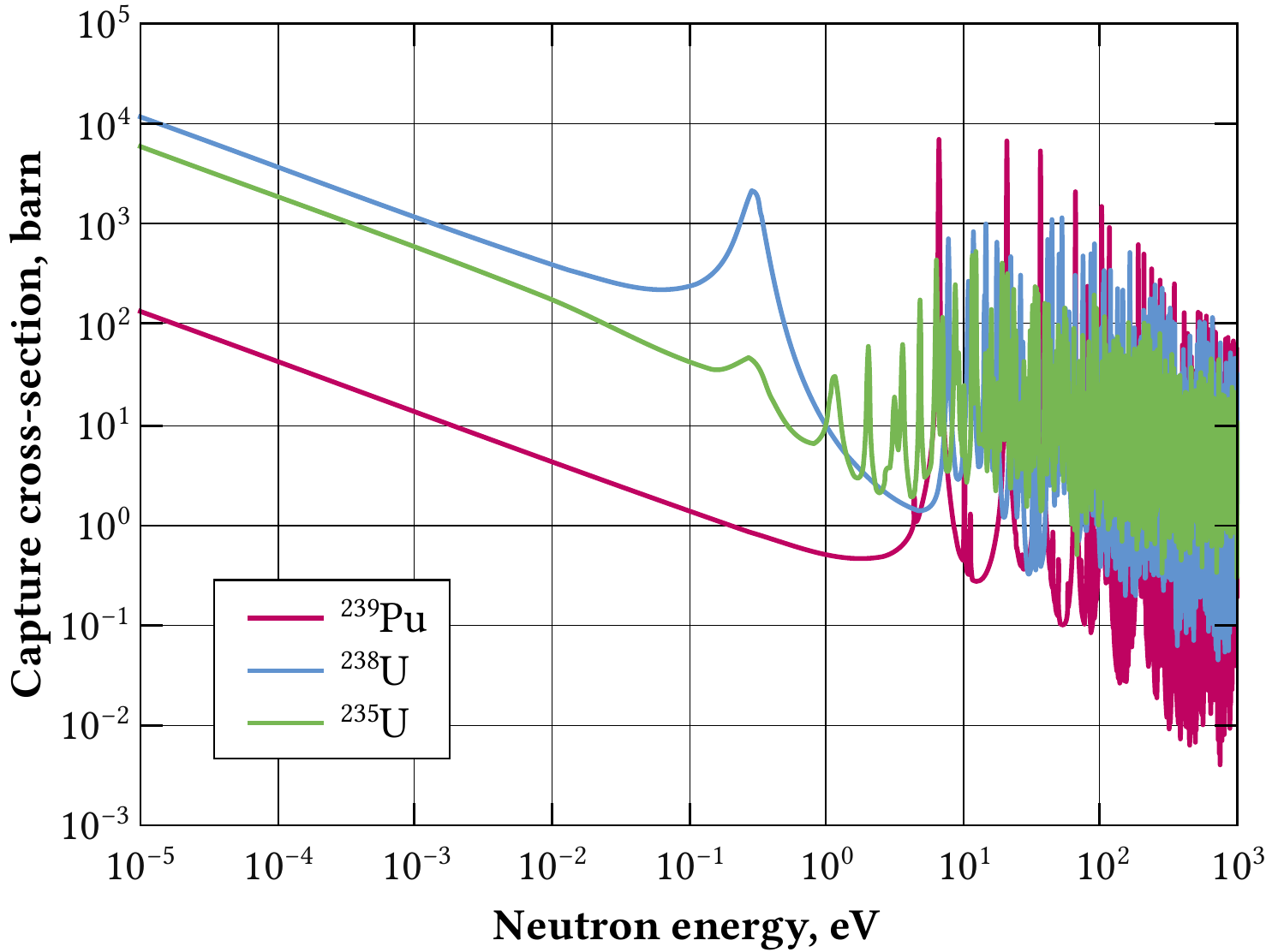}
\caption{The dependences of radiation capture cross-sections on the energy of 
         neutrons for $^{239}Pu$, $^{235}U$ and $^{238}U$, taken from the 
	 database ENDF/B-VII.0.}
\label{fig:02}
\end{figure}

In the figures~\ref{fig:01} and~\ref{fig:02} the dependences of the 
cross-sections of nuclear reactions of fission and radiative capture on the 
energy of neutrons are given. In these figures the energy of neutrons ranges 
from 10$^{-5}$~eV to 10$^7$~eV.

\begin{figure}[htb]
\centering
\includegraphics[width=16cm]{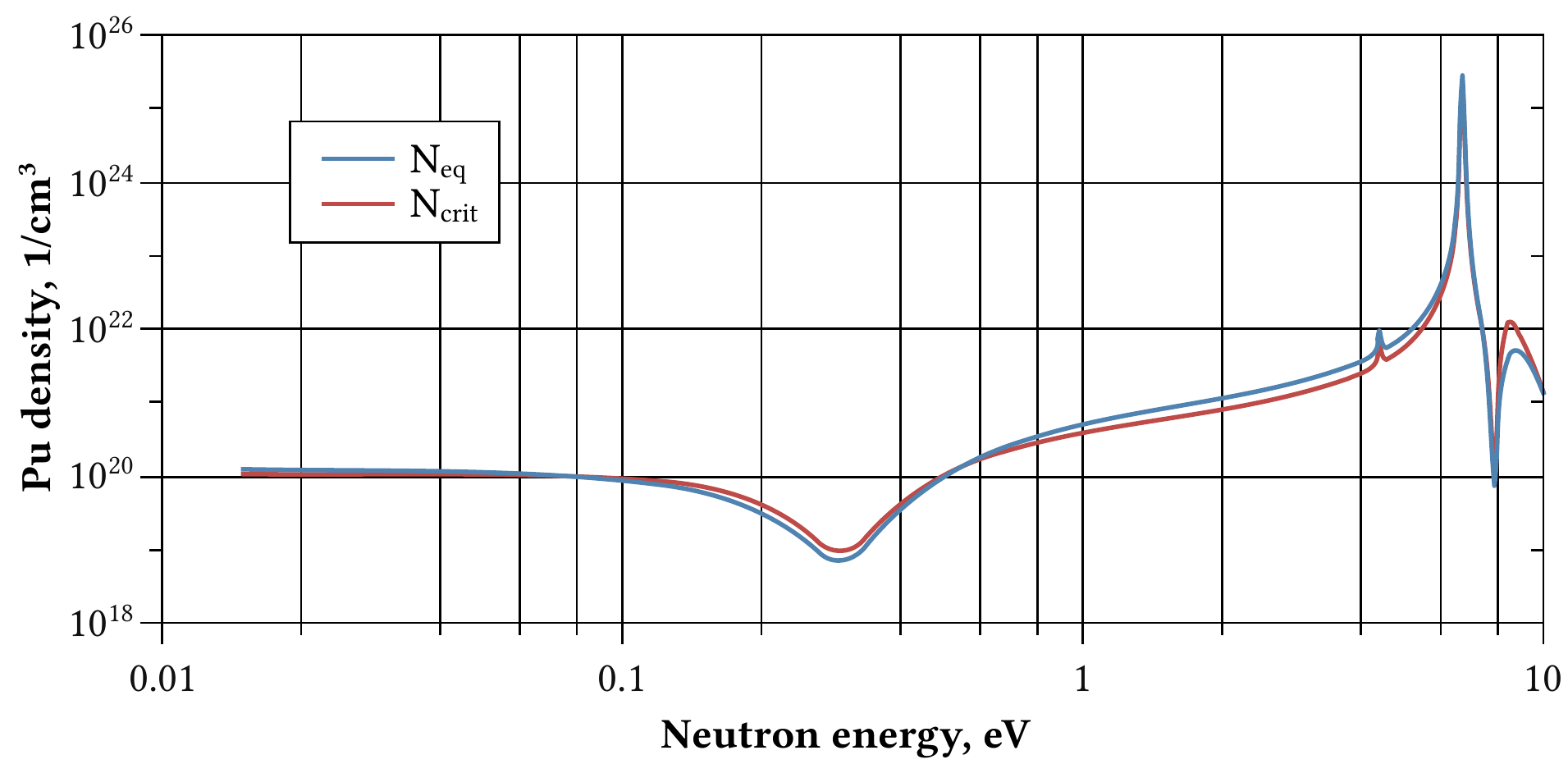}
\caption{The simulation of dependences of equilibrium $N_{eq}^{Pu}$ and critical
         $N_{crit}^{Pu}$ densities of $^{239}Pu$ on the energy of neutrons, 
	 ranging from 0.015~eV to 10.05~eV.}
\label{fig:03}
\end{figure}

\begin{figure}[htb]
\centering
\includegraphics[width=16cm]{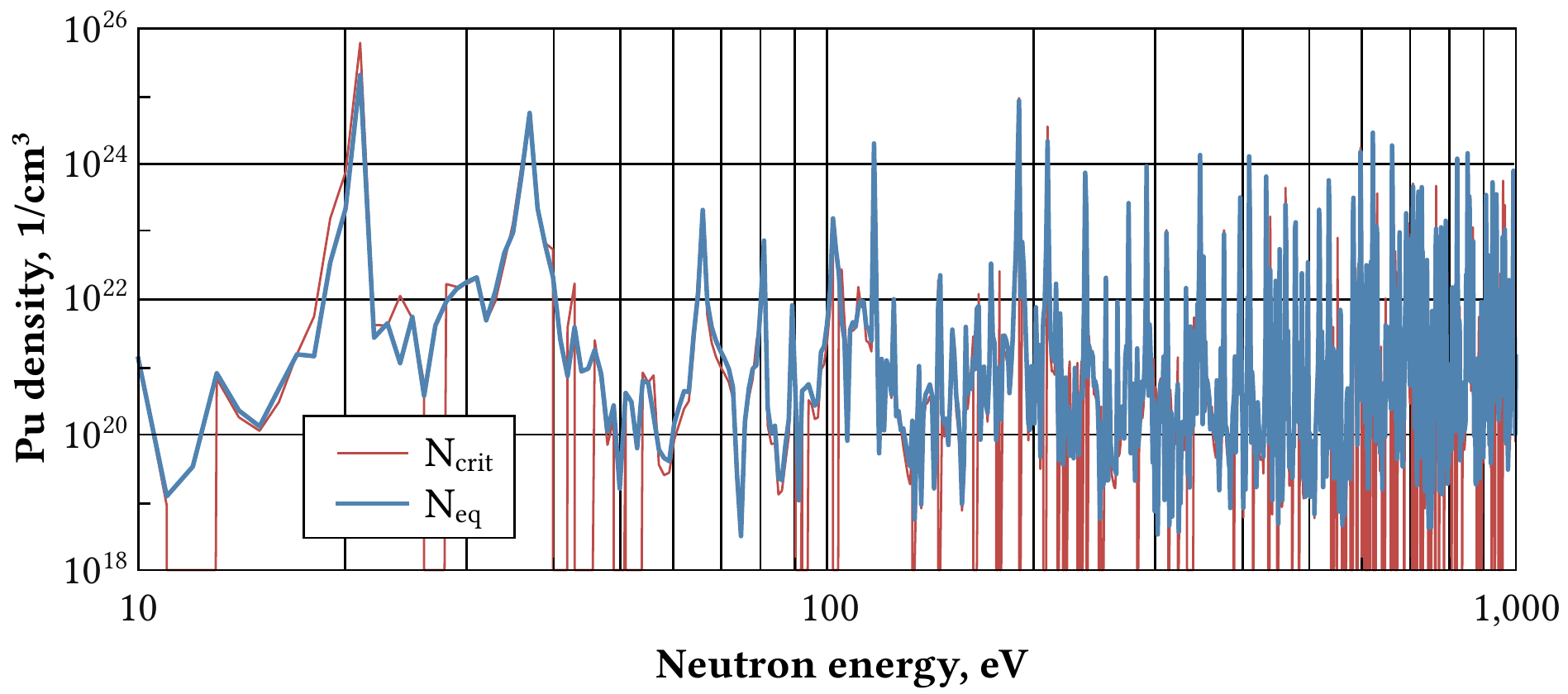}
\caption{The simulation of dependences of equilibrium $N_{eq}^{Pu}$ and critical
         $N_{crit}^{Pu}$ densities of $^{239}Pu$ on the energy of neutrons, 
	 ranging from 10~eV to 1~keV.}
\label{fig:04}
\end{figure}

\begin{figure}[htb]
\centering
\includegraphics[width=16cm]{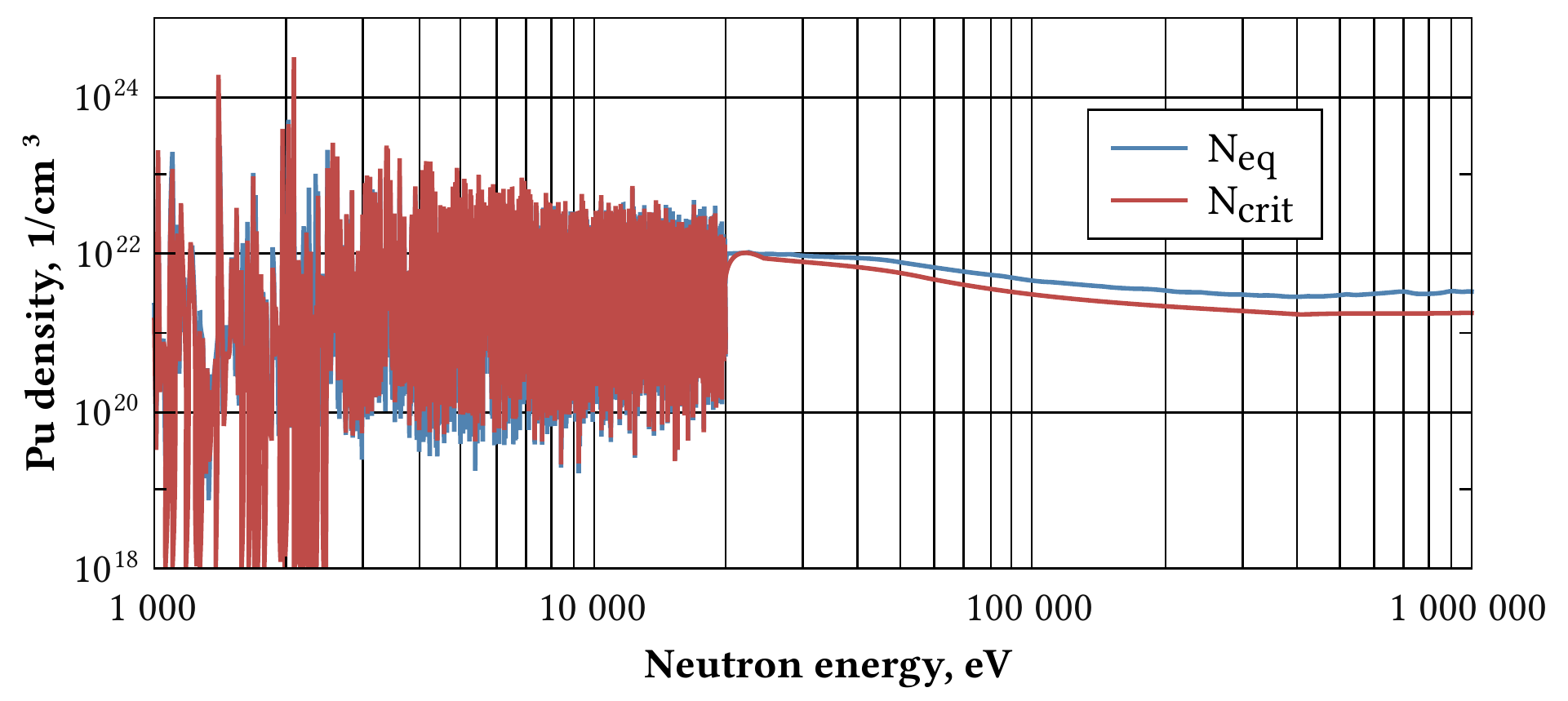}
\caption{The simulation of dependences of equilibrium $N_{eq}^{Pu}$ and critical
         $N_{crit}^{Pu}$ densities of $^{239}Pu$ on the energy of neutrons, 
         ranging from 1~keV to 1~MeV.}
\label{fig:05}
\end{figure}

According to the relationships~(\ref{eq21}) and~(\ref{eq22}), we have calculated
the equilibrium $N_{eq}^{Pu}$ and critical $N_{crit}^{Pu}$ concentrations of 
$^{239}Pu$ for the uranium-plutonium fissile medium. The results of the 
calculations are presented in figures~\ref{fig:03}~–~\ref{fig:05}.

The analysis of the presented in figures~\ref{fig:03}~–~\ref{fig:05} results 
allows to draw a conclusion that there exist several regions of energies of 
neutrons 0.015$\div$0.05~eV (Fig.~\ref{fig:03}), 0.6$\div$6~eV 
(Fig.~\ref{fig:03}), 90$\div$300~eV (Fig.~\ref{fig:04}) and 0.24$\div$1~MeV 
(Fig.~\ref{fig:05}), where the Feoktistov criterion 
$N_{eq}^{^{239}_{94}Pu} > N_{crit}^{^{239}_{94}Pu}$ holds true, i.e. in these 
regions of energies realization of modes of wave neutron-nuclear burning is 
possible.

Thus, in contrast to the conclusion about a possibility of wave neutron-nuclear 
burning only in the region of fast neutrons, drawn in~\cite{ref01} and based on 
estimates of the equilibrium $N_{eq}^{Pu}$ and critical $N_{crit}^{Pu}$ 
concentrations of $^{239}Pu$ only for two values of the neutron energy (thermal 
0.025~eV, fast 1~MeV), in the present paper we find out the fulfilment of the 
Feoktistov criterion and, consequently, a possibility of realization of modes of
neutron-nuclear burning also in the region of cold, epithermal and resonance 
neutron energies.

Let us consider the thermal region of neutron energies. Existence of regions of 
neutron-nuclear burning for the neutron energies 0.015$\div$0.05~eV and 
0.6$\div$6~eV, in contrast to the region of energies 0.05$\div$0.6~eV, where the
Feoktistov criterion does not hold true (see Fig.~\ref{fig:03}) and, 
consequently, the mode of neutron-nuclear burning is not realized, can be 
explained by the presence of the resonance on the curve of the $^{239}Pu$ 
radiation capture cross-section dependence on the neutron energy in the range of
energies 0.05$\div$0.6~eV (Fig.~\ref{fig:02}) and the analytical form of the 
expressions~(\ref{eq21}) and~(\ref{eq22}) for the equilibrium and critical 
concentrations of $^{239}Pu$. Indeed, the cross-section of the neutron radiative
capture for $^{239}Pu$ is a positive term in the denominator of the 
expression~(\ref{eq21}) for the equilibrium concentration of $^{239}Pu$, that 
leads to the abrupt decrease of the $^{239}Pu$ equilibrium concentration in the 
resonance energy range 0.05$\div$0.6~eV on the curve of the $^{239}Pu$ radiation
capture cross-section dependence. In contrast to the expression~(\ref{eq21}), 
the neutron radiation capture cross-section for $^{239}Pu$ is a negative term in
the denominator of the expression~(\ref{eq22}) for the critical concentration of
$^{239}Pu$, which leads to the abrupt increase of the $^{239}Pu$ critical 
concentration in the resonance energy range 0.05$\div$0.6~eV on the curve of the
$^{239}Pu$ radiation capture cross-section dependence.

In the region of resonance neutron energies 90$\div$300~eV the visual analysis 
of the results, presented in Fig.~\ref{fig:04}, discovers the region of possible
realization of the neutron-nuclear burning mode. Let us also note that for this 
resonance energy region it is possible to make more correct estimation of the 
equilibrium and critical concentrations of $^{239}Pu$ as a result of averaging 
over the neutron energy spectrum.

At the same time in the region of fast neutrons 0.24$\div$1~MeV 
(Fig.~\ref{fig:05}), similar to~\cite{ref01}, a possibility of the 
neutron-nuclear burning mode is confirmed.

\section{Generalized Feoktistov's criterion for Uranium-Plutonium neutron
         multiplication medium}

Let us note that in~(\ref{eq21}) and~(\ref{eq22}) the arguments, on which the 
concentrations of the nuclides must depend, are not shown explicitly, since we 
consider kinetics of the system of neutrons and nuclides, and the arguments, on 
which the cross-sections depend, are also not shown, and therefore one has the 
impression that $N_{eq}^{Pu}$ and $N_{crit}^{Pu}$ are constants. However, this 
simplification has reasonable explanation. In~\cite{ref01} the idea of a 
possibility of the nuclear burning wave existence itself was grounded, and at 
least approximate estimates, confirming this at least for particular cases of 
equilibrium $N_{eq}^{Pu}$ and critical $N_{crit}^{Pu}$ concentrations of 
$^{239}Pu$ were necessary. The author of~\cite{ref01}, apparently, reasoned in 
the following way: since the concentration of uranium-238 in the local region of
the fissile medium for the system considered in~\cite{ref01}, being the closest 
to the external neutron source, only decreases from the initial concentration 
100\% with the lapse of time, the maximum value of the estimate of the 
equilibrium concentration of plutonium-239, according to~(\ref{eq21}) (where the
cross-sections are constants for the fixed neutron energy), must be obtained 
exactly for this initial maximum concentration of uranium-238. At the same time 
the concentration of plutonium-239 in the same local region grows from zero to 
its maximum value with the lapse of time. As mentioned above, according to the 
Feoktistov criterion, the relationship 
$N_{eq}^{^{239}_{94}Pu} > N_{crit}^{^{239}_{94}Pu}$ must hold true for the 
appearance of the nuclear burning wave, therefore, if this relationship does not
hold true for the initial concentration of uranium-238, it will not hold true 
according to~(\ref{eq21}) later as well, and in such a system a nuclear burning 
wave should not exist. Obviously, therefore in~\cite{ref01} with the help of the
expressions~(\ref{eq21}) and~(\ref{eq22}) for the fixed concentration of 
uranium-238, being equal to 100\% (engineering uranium), for two fixed neutron 
energies (1~MeV for fast and 0.025~eV for thermal neutrons) estimations were 
made, demonstrating the fulfilment of the Feoktistov criterion for fast neutrons
and, consequently, also a possibility of existence of a nuclear burning wave for
a fast uranium-plutonium reactor and its impossibility for a slow 
uranium-plutonium reactor.

However, as said above, everything is much more complicated in reality.

Taking into account the real neutron spectrum in the fissile medium and 
introducing the probability function of the neutron distribution over the energy
spectrum $\rho (r,E_n,t)$, for this function we can write down the formula

\begin{equation}
\rho (\vec{r},E_n,t) = \frac{\Phi (\vec{r},E_n,t)}{\int \Phi (\vec{r},E_n,t) dE_n} = \frac{\Phi (\vec{r},E_n,t)}{\Phi(\vec{r},t)},
\label{eq23}
\end{equation}

\noindent where the total neutron density flux reads 
$\Phi (\vec{r},t) = \int \Phi (\vec{r},E_n,t) dE_n$.

Further, if one takes into account the three-dimensional geometry, the 
dependence of cross-sections on the neutron energy $E_n$ and the temperature of 
the fissile medium $T$ (taking into account the temperature influence is 
obligatory, since the wave nuclear burning modes with burning down to 50\%~\cite{ref01,ref02,ref03,ref04,ref05} can be realized) by preservation of the 
simplifying assumptions (as in~\cite{ref01}, the kinetic equation for 
plutonium-239 is written assuming that uranium-238 turns directly into 
plutonium-239 with some typical transition time, the delayed neutrons are 
disregarded), which allow to preserve the general form of the 
expressions~(\ref{eq21}) and~(\ref{eq22}), we can rewrite them in the new form. 
In order to do it, we have to return to the balance equations for plutonium-239 
and neutrons, similar to the same equations in~\cite{ref01}, and from which 
in~\cite{ref01} the expressions~(\ref{eq21}) and~(\ref{eq22}) were obtained:

\begin{equation}
\frac{\partial N^{Pu} (\vec{r},T,E_n,t)}{\partial t} \approx \Phi (\vec{r},E_n,T,t) \left[ \sigma_a ^{8} N^8 (\vec{r},T,t) - \sigma_a ^{Pu} (E_n,T) N^{Pu} (\vec{r},T,t) \right]
\label{eq24}
\end{equation}

\noindent and

\begin{equation}
\frac{\partial n (\vec{r},T,E_n,t)}{\partial t} \approx \Phi (\vec{r},E_n,T,t) 
\left[ \sum \limits_i \nu_i \sigma_f ^{i}(E_n,T) N^i (\vec{r},T,t) - \sum \limits_i \sigma_a ^{i} (E_n,T) N^{i} (\vec{r},T,t) \right]
\label{eq25}
\end{equation}

\noindent
where $n(\vec{r},E_n,T,t)$ and $N^{Pu} (\vec{r},T,E_n,t)$ represent phase 
concentrations of neutrons and plutonium, respectively, while $\Phi (\vec{r},E_n,T,t)$ is the phase neutron density flux.

Integrating left and right hand sides of the expressions~(\ref{eq24}) 
and~(\ref{eq25}) over the neutron energy and dividing them by the total neutron density flux $\Phi (\vec{r},t)$, for the total plutonium concentration $N^{Pu} (\vec{r},T,t)$ and the total neutron density $n(\vec{r},T,t)$, taking into 
account~(\ref{eq23}), we obtain the following expressions:

\begin{equation}
\frac{\partial N^{Pu} (\vec{r},T,t)}{\partial t} \approx 
\Phi (\vec{r},T,t) \left[ \bar{\sigma}_a ^{8} N^8 (\vec{r},T,t) - \bar{\sigma}_a ^{Pu} (T) N^{Pu} (\vec{r},T,t) \right]
\label{eq26}
\end{equation}

\noindent and

\begin{equation}
\frac{\partial n (\vec{r},T,t)}{\partial t} \approx \Phi (\vec{r},T,t) 
\left[ \sum \limits_i \nu_i \bar{\sigma}_f ^{i}(T) N^i (\vec{r},T,t) - \sum \limits_i \bar{\sigma}_a ^{i} (T) N^{i} (\vec{r},T,t) \right]
\label{eq27}
\end{equation}

\noindent
where $\bar{\sigma} _j ^i (\vec{r},T,t) = \int \sigma _j ^i (E_n,T) \rho (\vec{r},E_n,t) dE_n$ are the cross-sections of the $j^{th}$ nuclear reaction for the 
$i^{th}$ nuclide of the fissile medium averaged over the neutron energy 
spectrum.

Further, from~(\ref{eq26}) and~(\ref{eq27}), equating the derivatives to zero, 
we obtain the following expressions for the equilibrium concentration 
$\tilde{N}_{eq}^{Pu}$ (of course, it is not an equilibrium, but some stationary 
concentration, but we use this term after Feoktistov) of the fissile nuclide 
$^{239}_{94}Pu$ and its critical concentration $\tilde{N}_{crit}^{Pu}$:

\begin{equation}
\tilde{N}_{eq}^{Pu} (\vec{r},T) \approx \frac{\bar{\sigma}_c ^8 (T)}{\bar{\sigma}_a ^{Pu} (T)} N^8 (\vec{r},T) 
\label{eq28}
\end{equation}

\noindent and

\begin{equation}
\tilde{N}_{crit}^{Pu} (\vec{r},T) \approx \frac{\sum \limits _{i \neq Pu} \bar{\sigma}_a ^i (T) N^i (\vec{r},T) - \sum \limits _{i \neq Pu} \nu_i \bar{\sigma}_f ^i (T) N^i (\vec{r},T)}{(\nu_{Pu}-1) \bar{\sigma}_f ^{Pu} (T) - \bar{\sigma}_c ^{Pu} (T)}
\label{eq29}
\end{equation}

Thus, keeping the physical meaning of the Feoktistov criterion, we will 
approximate to the more realistic analysis of a possibility of a nuclear burning
wave realization in the uranium-plutonium fissile medium, if we rely on the 
relationship

\begin{equation}
\tilde{N}_{eq}^{^{239}_{94}Pu} > \tilde{N}_{crit}^{^{239}_{94}Pu}
\label{eq30}
\end{equation}

\noindent
where $\tilde{N}_{eq}^{^{239}_{94}Pu}$ and $\tilde{N}_{crit}^{^{239}_{94}Pu}$ 
are set by the expressions~(\ref{eq28}) and~(\ref{eq29}).

For the calculation of the cross-sections of the $j^{th}$ nuclear reaction for 
the $i^{th}$ nuclide of the fissile medium $\bar{\sigma}_j ^i (\vec{r},T,t)$ 
averaged over the neutron energy spectrum, present in the 
expressions~(\ref{eq28}) and~(\ref{eq29}), on which, consequently, the 
fulfilment of the Feoktistov criterion~(\ref{eq30}) depends, one needs to know 
(be able to calculate) the energy spectrum of the slowing-down neutrons and the 
dependences of the nuclear reactions cross-sections on the neutron energy and 
the temperature of the fissile medium (the Doppler effect).

In order to calculate the cross-sections averaged over the neutron energy 
spectrum we used the spectrum of thermal neutrons of the WWER 
reactor~\cite{ref16}.

\begin{figure}[htb]
\centering
\includegraphics[width=16cm]{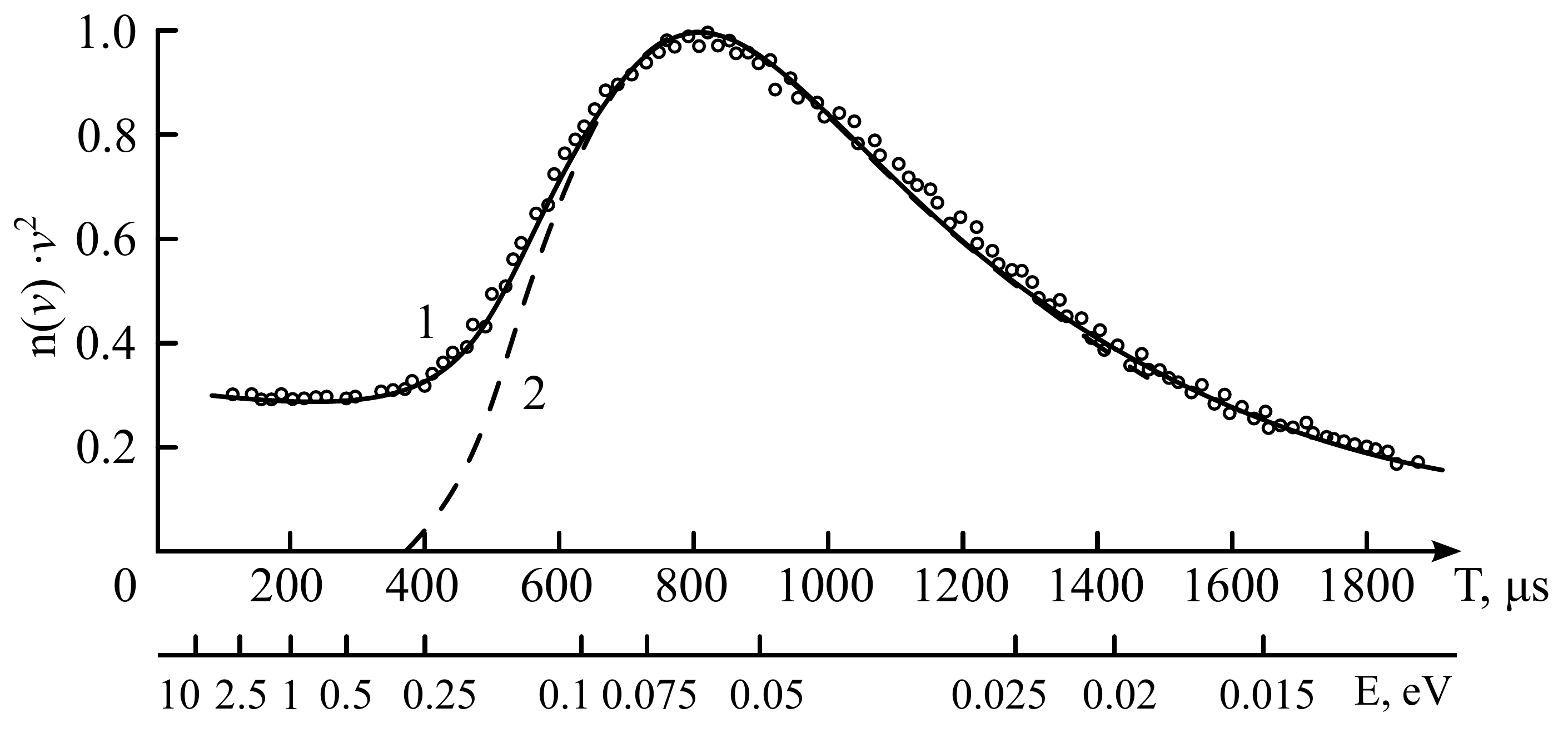}
\caption{Energy spectrum of neutrons in the WWER~\cite{ref16} (1 – spectrum of 
         neutrons; 2 – Maxwellian distribution; $T$ is a transit time (in 
	 microseconds) for some standard distance; $E$ is the energy of 
	 neutrons, corresponding to this transit time; $n(v)$ is the density of
	 neutrons with the velocity $v$).}
\label{fig:06}
\end{figure}

The equilibrium and critical concentrations of $^{239}Pu$ were calculated with 
the help of the expressions~(\ref{eq28}) and~(\ref{eq29}), into which the 
cross-sections of nuclear reactions averaged over the neutron spectrum of the 
WWER enter. The following estimates of the averaged cross-sections of fission 
and radiation capture for uranium-238 and plutonium-239 were obtained:

\begin{equation}
\bar{\sigma} _c ^{Pu} = 339.10 b, ~ ~ \bar{\sigma} _f ^{Pu} = 553.60 b, ~ ~
\bar{\sigma} _c ^{238} = 255.33 b, ~ ~ \bar{\sigma} _f ^{238} = 0.00 b.
\end{equation}

Using the obtained estimates for the averaged cross-sections for nuclides of 
uranium-238 and plutonium-239 according to the expressions~(\ref{eq08}) 
and~(\ref{eq09}) we calculated the equilibrium and critical concentrations of 
$^{239}Pu$ and obtained the following values:

\begin{equation}
\tilde{N}_{eq}^{Pu} \approx 1.302 \cdot 10^{22} cm^{-3} ~ ~ and ~ ~
\tilde{N}_{crit}^{Pu} \approx 1.618 \cdot 10^{22} cm^{-3}.
\end{equation}

The obtained estimates show that the Feoktistov's criterion does not hold true. 
This can be explained by the fact that the majority of neutrons of thermal WWER 
reactor, according to Fig.~\ref{fig:06}, corresponds to the energy range 
0.05$\div$0.6~eV, coinciding with the resonance energy range on the curve of the
$^{239}Pu$ radiative capture cross-section dependence (Fig.~\ref{fig:02}), that,
as we saw in the previous section, leads to the decrease of the equilibrium 
concentration of $^{239}Pu$ and to the increase of the critical one.

Also we calculated the values of averaged cross-sections of fission and 
radiation capture for uranium-238 and plutonium-239 as well as the equilibrium 
and critical concentrations of $^{239}Pu$ (the expressions~(\ref{eq28}) 
and~(\ref{eq29})) for the neutron energy ranges 0.015$\div$0.05~eV and 
0.60$\div$6.00~eV of the WWER reactor spectrum, in which, as we saw in 
section~\ref{sec02} (see Fig.~\ref{fig:03}), in concordance with the obtained 
earlier estimates the Feoktistov criterion holds true. The following estimates 
were obtained:

\begin{itemize}
\item for the neutron energy range 0.015$\div$0.06~eV

$\bar{\sigma} _c ^{Pu} = 5.691 b$, $\bar{\sigma} _f ^{Pu} = 14.821 b$,
$\bar{\sigma} _c ^{238} = 0.052 b$, $\bar{\sigma} _f ^{238} = 0.00 b$;

$\tilde{N}_{eq}^{Pu} \approx 1.152 \cdot 10^{20} cm^{-3}$ and 
$\tilde{N}_{crit}^{Pu} \approx 1.045 \cdot 10^{20} cm^{-3}$;

\item for the neutron energy range 0.60$\div$6.00~eV

$\bar{\sigma} _c ^{Pu} = 13.161 b$, $\bar{\sigma} _f ^{Pu} = 48.199 b$,
$\bar{\sigma} _c ^{238} = 2.199 b$, $\bar{\sigma} _f ^{238} = 0.00 b$;

$\tilde{N}_{eq}^{Pu} \approx 1.632 \cdot 10^{21} cm^{-3}$ and 
$\tilde{N}_{crit}^{Pu} \approx 1.269 \cdot 10^{21} cm^{-3}$;

\end{itemize}

As it is seen from the obtained values for the equilibrium and critical 
concentrations of $^{239}Pu$, in these regions of energies of the WWER reactor 
spectrum the generalized over the spectrum criterion of realization of the 
neutron-nuclear wave burning mode holds true.

\section{Fulfilment of Feoktistov's criterion for fissile medium, originally 
         consisting of uranium-238 dioxide with enrichments 4.38\%, 2.00\%, 
	 1.00\%, 0.71\% and 0.50\% with respect to uranium-235, for the region 
	 of neutron energies 0.015$\div$10.00~eV}

\label{sec04}

Existence of three fissile nuclides of uranium-235, plutonium-239 and 
uranium-233, two latter of which are fissile nuclides primarily for ultraslow 
neutron-nuclear burning in uranium-plutonium and thorium-uranium nuclear 
reactions chains, respectively, allows to consider the realization of ultraslow 
neutron-nuclear burning modes more complicated with respect to the initial 
composition of the fissile media~\cite{ref04}, than those, which were considered
in the papers by Feoktistov and Teller. For example, one may consider a fissile 
medium originally consisting of uranium-238 and uranium-235 with different 
enrichments with respect to uranium-235, that corresponds to most widespread 
nuclear fuel of modern nuclear reactors, a fissile medium originally consisting 
of uranium-238 and plutonium-239 with different enrichments with respect to 
plutonium-239, the fissile medium originally consisting of thorium-232 and 
uranium-233 with different enrichments with respect to uranium-233, as well as 
a fissile medium originally consisting of any combination of those. It is clear 
that such broadening of possible compositions of fissile media enables to 
control the possible modes of ultraslow neutron-nuclear burning.

\begin{figure}[tb!]
\centering
\includegraphics[width=16cm]{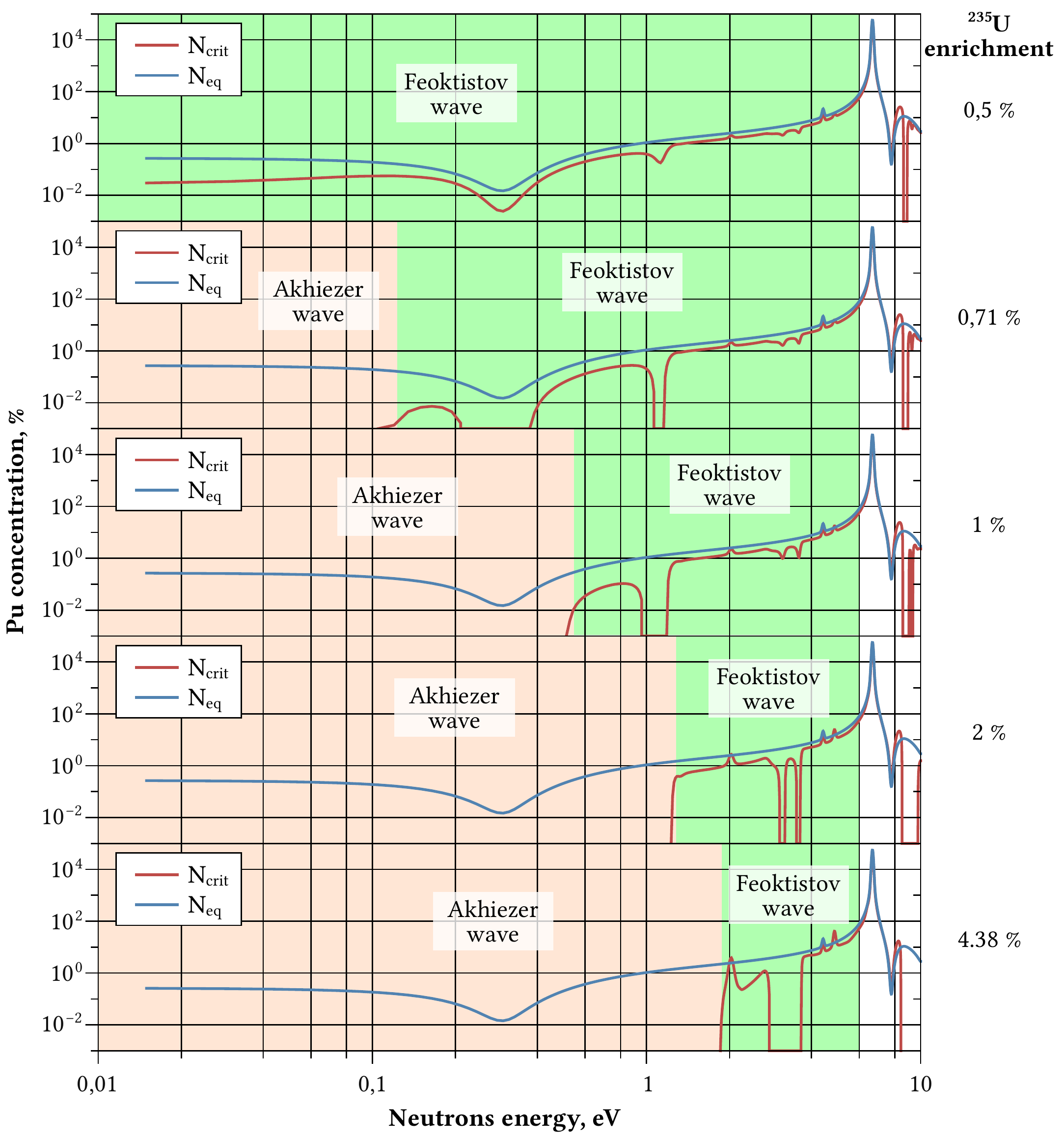}
\caption{The dependences of the equilibrium and critical concentrations of 
         $^{239}Pu$ on the neutron energy in the interval 0.015$\div$10~eV, 
	 originally consisting of uranium-238 dioxide with enrichments 4.38\%, 
	 2.00\%, 1.00\%, 0.71\% (natural uranium) and 0.50\% with respect to 
	 uranium-235.}
\label{fig:07}
\end{figure}

For the fissile medium consisting of uranium-238 dioxide with enrichments 
4.38\%, 2.00\%, 1.00\%, 0.71\% (natural uranium) and 0.5\% with respect to 
uranium-235, for the neutron energy region 0.015$\div$10.00~eV according to the 
expressions~(\ref{eq21}) and~(\ref{eq22}) we calculated the equilibrium and 
critical concentrations of $^{239}Pu$. The results of calculations are 
presented in Fig.~\ref{fig:07}.

The results presented in Fig.~\ref{fig:07} indicate that, e.g. for the fissile 
medium, consisting of uranium-238 dioxide with the enrichment 4.38\% with 
respect to uranium-235, the critical concentration of plutonium-239 exceeds zero
only for the neutron energy range 1.8$\div$10.0~eV (for the neutron energies 
less than 1.8~eV the uranium fissile medium with the enrichment 4.38\% with 
respect to uranium-235 is already in the above-critical state; it is well known 
that the natural uranium under the thermal energy of neutrons 0.025~eV has the 
criticality factor 1.32 (see e.g.~\cite{ref17}), and therefore the critical 
concentration of plutonium-239 in this neutron energy region, calculated by the 
formula~(\ref{eq22}), is negative and is not presented in Fig.~\ref{fig:07}). 
Practically in the whole this region of neutron energies 1.8$\div$6.0~eV the 
Feoktistov's criterion holds true, and so does the generalized over the neutron 
spectrum criterion. Thus, if one forms such composition, structure and geometry 
of an active zone of a nuclear reactor that the neutron spectrum corresponds 
mainly to this neutron energy region, then realization of such wave nuclear 
reactor is possible.

Similar conclusions are true also for all other dependences, presented in 
Fig.~\ref{fig:07}.

Let us also note that the equilibrium concentration, calculated by means of the 
approximate expression~(\ref{eq21}), does not take into account the possibility 
of the initial enrichment of the uranium-238 dioxide by uranium-235 and 
therefore does not change under different enrichments with respect to 
uranium-235.

It is important to note that the results presented in Fig.~\ref{fig:07} 
demonstrate the division of the considered neutron energy region into energy 
regions, where the Akhiezer mode of slow burning is realized (the regions, where
the estimate of the critical concentration of plutonium-239 is negative, e.g. 
as already mentioned above for natural uranium (the enrichment 0.71\% with 
respect to uranium-235), the region of neutron energies less than 1.8~eV) and 
energy regions, where the Feoktistov mode of ultraslow burning is realized (the 
regions, where the estimate of the critical concentration of plutonium-239 is 
positive and the Feoktistov criterion holds true, e.g. as already mentioned 
above for the natural uranium, the region of neutron energies less than 
1.8$\div$6.0~eV) for the given compositions of the fissile medium. In 
Fig.~\ref{fig:07} those regions, where the modes of slow Akhiezer burning and 
ultraslow Feoktistov burning are realized, are coloured differently for clearness.

Thus, the general criterion of the wave neutron-nuclear burning modes 
realization (both Akhiezer and Feoktistov waves) could be formulated in the 
following way:

\begin{itemize}
\item if the neutron-multiplication fissile medium is originally (before the 
      action of the external source of neutrons) in the above-critical state, 
      then under the action of the external source of neutrons the Akhiezer wave
      of slow burning is realized in it;

\item if the neutron-multiplication fissile medium is originally in the critical
      state and if the Feoktistov criterion holds true, then under the action of
      the external source of neutrons the Feoktistov wave of ultraslow burning 
      is realized in it.
\end{itemize}

The analysis of the results presented in Fig.~\ref{fig:07} also allows to 
conclude that when the enrichment of uranium-238 dioxide decreases from 4.38\% 
to 0.50\% with respect to uranium-235, the broadening of the neutron energy 
region, where the critical concentration of plutonium-239 exceeds zero, to 
thermal and even cold neutron energies happens, and in all these regions the 
Feoktistov criterion $N_{eq}^{^{239}_{94}Pu} > N_{crit}^{^{239}_{94}Pu}$ and the generalized over the neutron spectrum criterion $\tilde{N}_{eq}^{^{239}_{94}Pu}
> \tilde{N}_{crit}^{^{239}_{94}Pu}$ (see~(\ref{eq30})) hold true, i.e. 
realization of the slow wave neutron-nuclear burning mode is possible.

For practical realization of wave reactors it is important to note that, as it 
follows from Fig.~\ref{fig:07}, in natural uranium dioxide (the enrichment 
0.71\% with respect to uranium-235) the criteria of the slow wave 
neutron-nuclear burning realization hold true practically for the whole regions 
of thermal and epithermal neutrons.

Thus, in the present paper it is concluded for the first time that creating a 
thermal-epithermal wave nuclear reactor with natural uranium in its various 
forms as a fuel is possible, which is substantiated by the calculations results.

Indeed, all publications have been discussing only different variants of wave 
fast neutron reactors so far. In the closing paragraph of~\cite{ref01} a 
supposition about a possibility in principle of wave burning of plutonium-239 in
a heavy-water thermal natural uranium reactor is made. However, this supposition
is wrong. Really, as it follows from the results presented above 
(Fig.~\ref{fig:07}) and their analysis, in order for the wave burning in the 
thermal region of neutron energies to exist, the neutron energy region, where 
the critical concentration of plutonium-239 is positive, should exist. And we 
know that natural uranium fuel (the enrichment 0.71\% with respect to 
uranium-235) exactly in a heavy-water reactor already has supercriticality (as 
is known, the criticality factor for natural uranium for thermal neutrons 
amounts to 1.32 (see e.g.~\cite{ref17}) and already exceeds 1.00), that ensures 
its advantages over light-water reactors. Thus, as it follows from the 
above-stated in this section, in a heavy-water natural uranium reactor there is 
no region, where the critical plutonium concentration is positive, and 
consequently, the mode of slow wave neutron-nuclear burning of plutonium-239 
cannot be realized. The same is true for the natural uranium reactors with a gas
coolant. In these reactors slow wave nuclear burning is possible for fuel with 
so lesser enrichment with respect to uranium-235, that in the thermal energy 
region the plutonium-239 criticality region exists (the estimate of the critical
concentration of plutonium-239 is positive). For example, uranium-238 dioxide 
with the enrichment 0.50\% with respect to uranium-235 (see Fig.~\ref{fig:11}) or
with even lesser enrichment, or engineering uranium, or spent nuclear fuel, 
satisfying this condition, will do.

And light-water thermal reactors with natural uranium fuel will right do for 
slow wave burning of plutonium-239, since natural uranium in them, as is well 
known, is in the subcritical state, and exactly because of this the additional 
enrichment of fuel 2.0\%$\div$3.5\% with respect to uranium-235 is required for 
such reactors operation.

\section{Modeling of neutron-nuclear burning of natural uranium for the 
         epithermal region of neutron energies}

\label{sec05}

In order to confirm the validity of the afore-cited estimates and conclusions, 
based on the analysis of the slow wave neutron-nuclear burning criterion 
fulfilment depending on the neutron energy, the numerical modelling of 
neutron-nuclear burning in natural uranium within the epithermal region of 
neutron energy (0.1$\div$7.0~eV) was carried out.

We consider a semispace with respect to the coordinate $x$, filled with natural 
uranium (99.28\% of uranium-238 and 0.72\% of uranium-235), which is lighted 
from the open surface by the neutron source. For simplicity the diffusion 
one-group approximation is considered (the neutron energy 1~eV). Uranium-238, 
absorbing a neutron, turns into uranium-239, which then as a result of two 
$\beta$-decays with a typical time of the $\beta$-decay $\tau_\beta \sim$3~days 
turns into fission-active isotope of plutonium-239. As shown above in 
Section~\ref{sec04}, in such a medium a slow neutron-fission wave of 
plutonium-239 burning can arise.

Taking into account the delayed neutrons, kinetics of such a wave is described 
by a system of 20 partial differential equations with nonlinear feedbacks 
concerning 20 functions  $n(x,t)$, $N_5(x,t)$, $N_8(x,t)$, $N_9(x,t)$, 
$N_{Pu}(x,t)$, $\tilde{N}_i ^{(Pu)}(x,t)$, $\tilde{N}_i ^{(5)}(x,t)$, 
$\bar{N}_i ^{(Pu)}(x,t)$, $\bar{N}_i ^{(5)}(x,t)$ of two variables $x$ and $t$, 
which can be written down in the following form.

First let us write the kinetic equation for the density of neutrons:

\begin{equation}
\frac{\partial n (x,t)}{\partial t} = D \Delta n (x,t) + q(x,t),
\label{eq31}
\end{equation}

\noindent where the source volume density $q(x,t)$ reads

\begin{align}
q(x,t) & = [\nu^{(Pu)} (1 - p^{(Pu)}) - 1] \cdot n(x,t) \cdot V_n \cdot \sigma_f^{Pu} \cdot N_{Pu} (x,t) + \nonumber \\
{} & + [\nu^{(5)} (1 - p^{(5)}) - 1] \cdot n(x,t) \cdot V_n \cdot \sigma_f^{5} \cdot N_{5} (x,t) + \ln 2 \cdot \sum \limits _{i=1} ^6 \left[ \frac{\tilde{N}_i ^{(Pu)}}{T_{1/2}^{i (Pu)}} + \frac{\tilde{N}_i ^{(5)}}{T_{1/2}^{i (5)}} \right] - \nonumber \\
{} & - n(x,t) \cdot V_n \cdot \left[ \sum \limits _{5,8,9,Pu} \sigma_c ^i \cdot
N_i (x,t) + \sum \limits _{i=1} ^6 \left[ \sigma_c ^{i(Pu)} \cdot \tilde{N}_i ^{(Pu)} (x,t) + \sigma_c^{i (5)} \cdot \tilde{N}_{i}^{(5)} (x,t) \right] \right] - \nonumber \\
{} & - n(x,t) \cdot V_n \cdot \left[ \sigma _c ^{eff (Pu)} (x,t) \bar{N}^{(Pu)} + \sigma _c ^{eff (5)} \cdot \bar{N}^{(5)} + \sigma_c ^{eff} \cdot \bar{\bar{N}} (x,t) \right],
\label{eq32}
\end{align}

\noindent
where $n(x,t)$ is the density of neutrons; $D$ is the neutron diffusion 
coefficient; $V_n$ is the neutron velocity (E~=~3~eV, the one-group 
approximation); $\nu^{(Pu)}$ and $\nu^{(5)}$ represent mean numbers of 
instantaneous neutrons per one fission event for $^{239}Pu$ and $^{235}U$, 
respectively; $N_5$, $N_8$, $N_9$, $N_{Pu}$ are densities of $^{235}U$, 
$^{238}U$, $^{239}U$, $^{239}Pu$, respectively; $\tilde{N}_i^{(Pu)}$ and 
$\tilde{N}_i^{(5)}$ are densities of surplus neutron fragments of fission of 
nuclei $^{239}Pu$ and $^{235}U$, respectively; $\bar{N}_i ^{(Pu)}$ and 
$\bar{N}_i ^{(5)}$ are densities of all other fragments of fission of nuclei 
$^{239}Pu$ and $^{235}U$, respectively; $\bar{\bar{N}} (x,t)$ is the density of 
accumulated nuclei of ``slags''; $\sigma_c$ and $\sigma_f$ are 
micro-cross-sections of neutron radiative capture and nucleus fission reactions;
the parameters $p_i$ ($p = \sum \limits _{i=1} ^6 p_i$) and $T_{1/2}^i$, 
characterizing groups of delayed neutrons for main fuel fissile nuclides are 
known and given e.g. in~\cite{ref17,ref18}. Let us note that while deriving the 
equation for $q(x,t)$ for taking into account the delayed neutrons, the 
Akhiezer-Pomeranchuk method~\cite{ref19} was used.

The last terms in square brackets in the right hand side of~(\ref{eq32}) were 
set according to the method of averaged effective cross-section for 
``slags''~\cite{ref17}, e.g. for fission fragments of nuclei:

\begin{equation}
n(x,t) V_n \sum \limits _{i = fission~fragments} \sigma_c ^i \bar{N}_i (x,t) = 
n(x,t) V_n \sigma_c ^{eff} \bar{N}(x,t),
\label{eq33}
\end{equation}

\noindent
where $\sigma_c ^{eff}$ is some effective micro-cross-section of the radiation 
capture of neutrons for fragments.

The kinetic equations for $\bar{N}^{(Pu)} (x,t)$ and $\bar{N}^{(5)}$ have the 
following form:

\begin{equation}
\frac{\partial \bar{N}^{(Pu)} (x,t)}{\partial t} = 2 \left( 1 - \sum \limits_{i=1} ^{6} p_i ^{(Pu)} \right) \cdot n(x,t) \cdot V_n \cdot \sigma_f ^{Pu} \cdot N_{Pu} (x,t) - V_n \cdot n(x,t) \cdot \sigma_c ^{eff (Pu)} \cdot \bar{N}^{(Pu)} (x,t)
\label{eq34}
\end{equation}

\noindent and

\begin{equation}
\frac{\partial \bar{N}^{(5)} (x,t)}{\partial t} = 2 \left( 1 - \sum \limits_{i=1} ^{6} p_i ^{(5)} \right) \cdot n(x,t) \cdot V_n \cdot \sigma_f ^{5} \cdot N_{5} (x,t) - V_n \cdot n(x,t) \cdot \sigma_c ^{eff (5)} \cdot \bar{N}^{(5)} (x,t)
\label{eq35}
\end{equation}

Consequently, we obtain the following system of 20 kinetic equations:

\begin{equation}
\frac{\partial n(x,t)}{\partial t} = D \Delta n(x,t) + q(x,t),
\label{eq36}
\end{equation}

\noindent where $q(x,t)$ is given by the expression~(\ref{eq32});

\begin{equation}
\frac{\partial N_8 (x,t)}{\partial t} = - V_n n(x,t) \sigma_c^8 N_8 (x,t);
\label{eq37}
\end{equation}

\begin{equation}
\frac{\partial N_9 (x,t)}{\partial t} = V_n n(x,t) \left[ \sigma_c^8 N_8 (x,t) - \sigma_c^9 N_9 (x,t) \right] - \frac{1}{\tau_\beta} N_9 (x,t);
\label{eq38}
\end{equation}

\begin{equation}
\frac{\partial N_{Pu} (x,t)}{\partial t} = \frac{1}{\tau_\beta} N_9 (x,t) - 
V_n n(x,t) ( \sigma_f^{Pu} + \sigma_c^{Pu}) N_{Pu} (x,t);
\label{eq39}
\end{equation}

\begin{equation}
\frac{\partial N_5 (x,t)}{\partial t} = - V_n n(x,t) ( \sigma_f^5 + \sigma_c^5) N_5 (x,t);
\label{eq40}
\end{equation}

\begin{align}
\frac{\partial \tilde{N}_i ^{(Pu)} (x,t)}{\partial t} & = p_i ^{(Pu)} \cdot V_n \cdot n(x,t) \cdot \sigma_f ^{Pu} \cdot N_{Pu} (x,t) - \frac{\ln 2 \cdot \tilde{N}_i ^{(Pu)} (x,t)}{T_{1/2} ^{i(Pu)}} - \nonumber \\
{} & - V_n \cdot n(x,t) \cdot \sigma_c ^{eff (Pu)} \cdot \tilde{N}_i ^{(Pu)} (x,t), ~ ~ i = 1,6;
\label{eq41}
\end{align}

\begin{align}
\frac{\partial \tilde{N}_i ^{(5)} (x,t)}{\partial t} & = p_i ^{(5)} \cdot V_n 
\cdot n(x,t) \cdot \sigma_f ^{5} \cdot N_{5} (x,t) - \frac{\ln 2 \cdot 
\tilde{N}_i ^{(5)} (x,t)}{T_{1/2} ^{i(5)}} - \nonumber \\
{} & - V_n \cdot n(x,t) \cdot \sigma_c ^{eff (5)} \cdot \tilde{N}_i ^{(5)}(x,t),
~~ i = 1,6;
\label{eq42}
\end{align}

\begin{equation}
\frac{\partial \overline{N} ^{(Pu)} (x,t)}{\partial t} = 
2 \left( 1 - \sum \limits _{i=1} ^6 p_i ^{(Pu)} \right) \cdot V_n \cdot n(x,t) 
\cdot \sigma_f ^{Pu} \cdot N_{Pu} (x,t) - V_n \cdot n(x,t) 
\cdot \sigma_c ^{eff (Pu)} \cdot \overline{N} ^{(Pu)} (x,t);
\label{eq43}
\end{equation}

\begin{equation}
\frac{\partial \overline{N} ^{(5)} (x,t)}{\partial t} = 
2 \left( 1 - \sum \limits _{i=1} ^6 p_i ^{(5)} \right) \cdot V_n \cdot n(x,t) 
\cdot \sigma_f ^{5} \cdot N_{5} (x,t) - V_n \cdot n(x,t) 
\cdot \sigma_c ^{eff (5)} \cdot \overline{N} ^{(5)} (x,t);
\label{eq44}
\end{equation}

\begin{align}
\overline{\overline{N}} (x,t) & = V_n n(x,t) \left[ \sigma_c ^9 N_9 (x,t) +
\sigma_c ^{Pu} N_{Pu} (x,t) + \sigma_c ^5 N_5 (x,t) + \phantom{\sum \limits_{i=1}^6} \right. \nonumber \\
{} & \left. + \sum \limits _{i=1} ^6 \left( \sigma_c ^{eff (Pu)} \tilde{N}_i ^{(Pu)} (x,t) + 
\sigma_c ^{eff (5)} \tilde{N}_i ^{(5)} (x,t) \right) + 
\sigma_c ^{eff (Pu)} \overline{N} ^{(Pu)} + 
\sigma_c^{eff (5)} \overline{N}^{(5)} (x,t) \right] + \nonumber \\
{} & + \sum \limits _{i=1} ^6 \left( \frac{\tilde{N}_i ^{(Pu)} \ln 2}{T_{1/2}^{i (Pu)}}
+ \frac{\tilde{N}_i ^{(5)} \ln 2}{T_{1/2} ^{i (5)}} \right) .
\label{eq45}
\end{align}

\noindent
where $\overline{\overline{N}} (x,t)$ is the total number of ``slagging'' 
nuclei, while $\tau_\beta$ is the lifetime of the nucleus with respect to the 
$\beta$-decay.

The boundary conditions:

\begin{equation}
\left. n(x,t) \right \vert _{x=0} = \frac{\Phi_0}{V_n} ~~~ \text{and} ~~~
\left. n(x,t) \right \vert _{x=l} = 0,
\label{eq46}
\end{equation}

\noindent
where $\Phi_0$ is the flux density of neutrons, created by a plane diffusion 
source of neutrons, situated at the border $x=0$; $l$ is the length of a block 
of natural uranium, set while modelling.

The initial conditions:

\begin{equation}
\left. n(x,t) \right \vert _{x=0,t=0} = \frac{\Phi_0}{V_n} ~~~ \text{and} ~~~
\left. n(x,t) \right \vert _{x \ne 0, t=0} = 0;
\label{eq47}
\end{equation}

\begin{equation}
\left. N_8(x,t) \right \vert _{t=0} = 0.9928 \cdot \frac{\rho_8}{\mu_8} N_A \approx 0.9928 \cdot \frac{19}{238} N_A  ~~~ \text{and} ~~~
\left. N_5(x,t) \right \vert _{t=0} \approx 0.7200 \cdot \frac{19}{238} N_A,
\label{eq48}
\end{equation}

\noindent
where $\rho _8$ is the mass density (g/cm$^3$) of uranium-238, $\mu_8$ is the 
molar mass ($g \cdot mole^{-1}$) of uranium-238, $N_A$ is the Avogadro number;

\begin{align}
& \left. N_9(x,t) \right \vert _{t=0} = 0, ~
\left. N_{Pu}(x,t) \right \vert _{t=0} = 0, ~
\left. \tilde{N}_i ^{(Pu)}(x,t) \right \vert _{t=0} = 0, \nonumber \\
& \left. \tilde{N}_i ^{(5)}(x,t) \right \vert _{t=0} = 0, ~
\left. \overline{N}_i^{(Pu)}(x,t) \right \vert _{t=0} = 0, ~
\left. \overline{N}_i^{(5)}(x,t) \right \vert _{t=0} = 0.
\label{eq49}
\end{align}

The numerical solution of the system of equations~(\ref{eq36}) – (\ref{eq45}) 
with the boundary and initial conditions~(\ref{eq46}) – (\ref{eq49}) was 
performed in Mathematica~8.

For optimization of the process of the numerical solution of the system of 
equations we passed to dimensionless quantities according to the following 
relationships:

\begin{equation}
n(x,t) = \frac{\Phi_0}{V_n} n^* (x,t), ~ ~
N(x,t) = \frac{\rho_8 N_A}{\mu_8} N^* (x,t).
\label{eq50}
\end{equation}

The model calculations were carried out for several variants of setting the constant coefficients of differential equations. Below in the present paper the results of two model calculations are presented.

For the first calculation the following numerical values were set for the constant coefficients of the differential equations:

\begin{align}
& D = 2.0 \cdot 10^4 ~cm^2 / s; ~
V_n = 1.0 \cdot 10^6 ~cm / s; ~
\Phi_0 = 1.0 \cdot 10^{23} ~cm^{-2} s^{-1}; ~
\tau_\beta \sim 3.3 ~days;  \nonumber \\
& \nu ^{(Pu)} = 2.90; \nu^{(5)} = 2.41; ~ \nonumber \\
& \sigma_f ^{Pu} = 477.04 \cdot 10^{-24} ~cm^2; ~
\sigma_c ^{Pu} = 286.15 \cdot 10^{-24} ~cm^2; ~
\sigma _c^8 = 252.50 \cdot 10^{-24} ~cm^2; ~ \nonumber \\
& \sigma_f ^5 = 136.43 \cdot 10^{-24} ~cm^2; ~
\sigma_c ^5 = 57.61 \cdot 10^{-24} ~cm^2; ~
\sigma_c ^9 = 4.80 \cdot 10^{-24} ~cm^2; ~ \nonumber \\
& \sigma_c ^{eff (Pu)} = 1.10 \cdot 10^{-24} ~cm^2; ~
\sigma_c ^{i (Pu)} = 1.00 \cdot 10^{-24} ~cm^2 , i=1..6; ~ \nonumber \\
& \sigma_c ^{eff (5)} = 1.10 \cdot 10^{-24} ~cm^2; ~
\sigma_c ^{i (5)} = 1.00 \cdot 10^{-24} ~cm^2, i=1..6; \nonumber \\
\nonumber \\
& T_1 ^{(Pu)} = 56.28 ~s; ~
T_2 ^{(Pu)} = 23.04 ~s; ~
T_3 ^{(Pu)} = 5.60 ~s; ~ \nonumber \\
& T_4 ^{(Pu)} = 2.13 ~s; ~
T_5 ^{(Pu)} = 0.62 ~s; ~
T_6 ^{(Pu)} = 0.26 ~s; ~ \nonumber \\
& p_1 ^{(Pu)} = 0.0072 \cdot 10^{-3}; ~
p_2 ^{(Pu)} = 0.626 \cdot 10^{-3}; ~
p_3 ^{(Pu)} = 0.444 \cdot 10^{-3}; ~ \nonumber \\
& p_4 ^{(Pu)} = 0.685 \cdot 10^{-3}; ~
p_5 ^{(Pu)} = 0.180 \cdot 10^{-3}; ~
p_6 ^{(Pu)} = 0.093 \cdot 10^{-3}; ~ \nonumber \\
& p^{(Pu)} = \sum \limits _{i=1} ^6 p_i ^{(Pu)} = 0.0021; ~
\sigma_c ^{eff} = 1.10 \cdot 10^{-24} ~cm^2; \nonumber \\
\nonumber \\
& T_1 ^{(5)} = 55.72 ~s; ~
T_2 ^{(5)} = 22.72 ~s; ~
T_3 ^{(5)} = 6.22 ~s; ~ \nonumber \\
& T_4 ^{(5)} = 2.30 ~s; ~
T_5 ^{(5)} = 0.61 ~s; ~
T_6 ^{(5)} = 0.23 ~s; ~ \nonumber \\
& p_1 ^{(5)} = 0.210 \cdot 10^{-3}; ~
p_2 ^{(5)} = 1.400 \cdot 10^{-3}; ~
p_3 ^{(5)} = 1.260 \cdot 10^{-3}; ~ \nonumber \\
& p_4 ^{(5)} = 2.520 \cdot 10^{-3}; ~
p_5 ^{(5)} = 0.740 \cdot 10^{-3}; ~
p_6 ^{(5)} = 0.27 \cdot 10^{-3}; ~ \nonumber \\
& p^{(5)} = \sum \limits _{i=1} ^6 p_i ^{(5)} = 0.0064; ~
l = 100 ~cm;
\label{eq51}
\end{align}

Let us note that the aforesaid cross-sections for the neutron-nuclear reactions for nuclides were set by their values averaged over the epithermal region of neutron energies (0.1$\div$7.0~eV).

During the calculation, the results of which are presented below in figures~\ref{fig:08}-\ref{fig:12}, the length of the fissile medium, where the wave of 
neutron-nuclear burning propagates, amounts to 100~cm, the full time of 
modelling is $t=30~min$, the temporal step is $\Delta t = 10 ~s$, the spatial 
coordinate step is $\Delta x = 0.01 ~cm$.

\begin{figure}[htb]
\centering
\includegraphics[width=4cm]{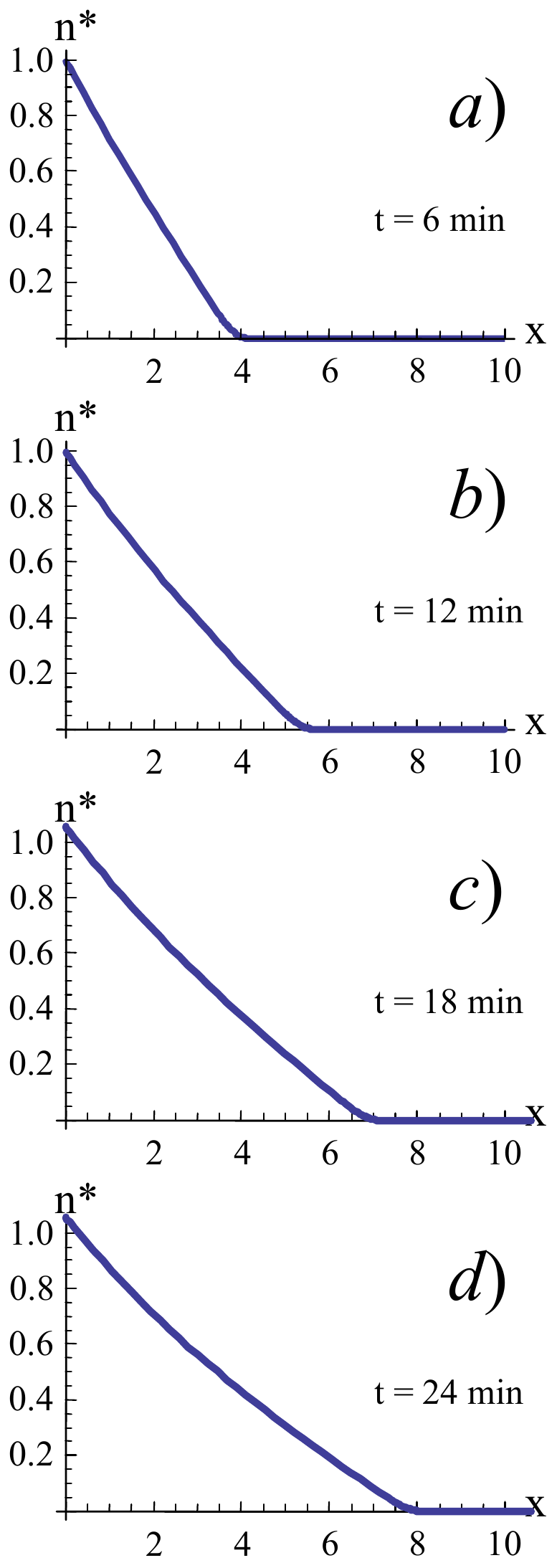}
\caption{Kinetics of neutrons under wave neutron-nuclear burning of natural 
         uranium. The dependence of the dimensionless density of neutrons on 
	 the spatial coordinate $n^* (x)$ for the moments of time 
	 (a) $t=6~min$; 
	 (b) $t = 12~min$; 
	 (c) $t = 18~min$; 
	 (d) $t = 24~min$).}
\label{fig:08}
\end{figure}

Of course, one would like to carry out the calculation for considerably longer 
computer experiment, to have the temporal step 
$\Delta t \approx 10^{-5} \div 10^{-7} ~s$ and to consider the external source 
of neutrons with the smaller flux density, but the authors of the paper, while 
choosing the aforesaid parameters for the calculation, were confined to their 
available computational resources.

The results of numerical modelling of the wave neutron-nuclear burning in 
natural uranium within the epithermal region of neutron energies 
(0.1$\div$7.0~eV) presented in figures~\ref{fig:08}–\ref{fig:12} indicate the 
realization of such a mode. Indeed in Fig.~\ref{fig:12} we can clearly see the 
wave burning of plutonium-239. At the same time, according to Fig.~\ref{fig:09} 
and Fig.~\ref{fig:10}, uranium-238 and uranium-235 burn down practically 
completely. It should be noted that the results for the neutron density kinetics
presented in Fig.~\ref{fig:08} do not demonstrate the neutron wave, in contrast 
to our results (e.g.~\cite{ref03,ref04,ref05,ref10}) published earlier, for 
neutron-nuclear burning of uranium-238 for fast neutrons (with the energy of the
order of 1~MeV). The authors explain this by the following fact. Since the 
system of differential equations was solved numerically for dimensionless 
(according to the relationships~(\ref{eq49})) variables, and while making them 
dimensionless, the density of neutrons was divided by the flux density of the 
external source, which was set by a specially overrated value 
$\Phi_0 = 1.0 \cdot 10^{23} ~cm^{-2} s^{-1}$ for the purpose of reducing the 
computation time, the difference between the scales of the external source flux 
density and the flux density of neutrons in the region of nuclear burning in the
steady-state regime of wave burning does not allow to see the neutron wave. It 
is also possible that the wave burning of plutonium-239 is not visible in 
Fig.~\ref{fig:08} for the density of neutrons, since burning of uranium-235 is 
superimposed on it. Really, the results of kinetics modelling for the 
uranium-235 nuclei density, presented in Fig.~\ref{fig:10}, show that 
uranium-235 burns down practically completely, and its concentration, being 
equal to 0.7\%, is almost three orders bigger than the amplitude of the steady 
wave concentration of plutonium-239, which according to Fig.~\ref{fig:12} equals
0.001\%.

\begin{figure}[htb]
\centering
\includegraphics[width=4cm]{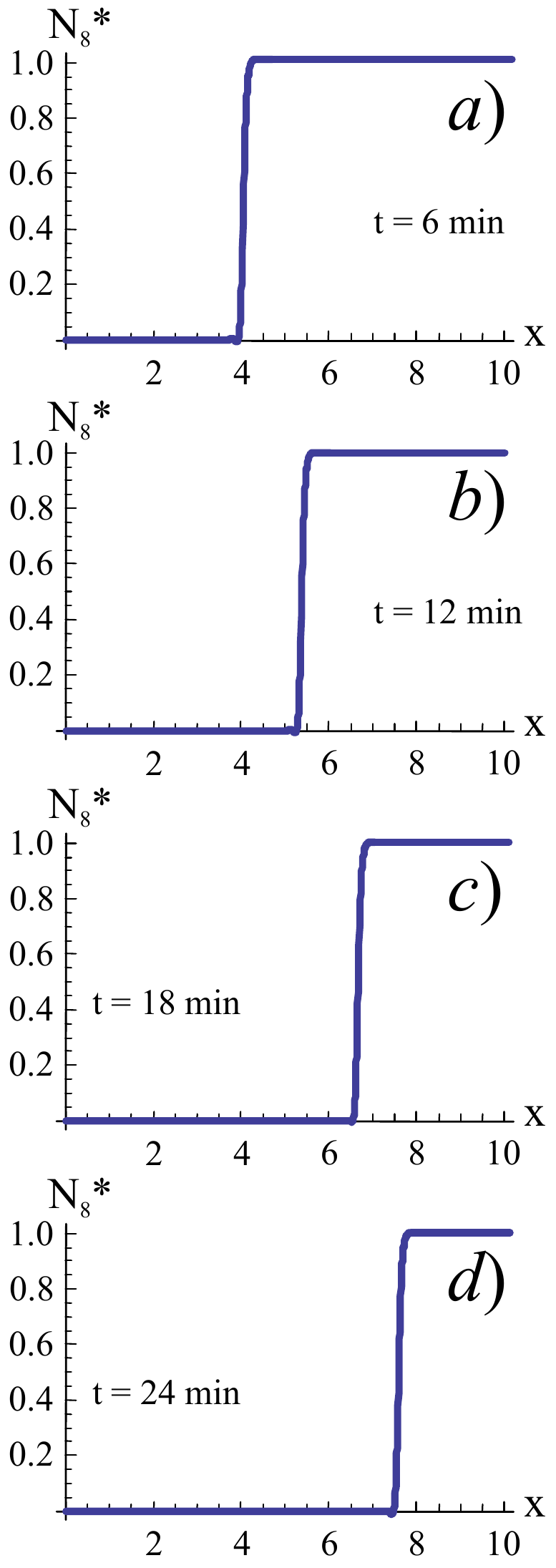}
\caption{Kinetics of the uranium-238 nuclei density under wave neutron-nuclear 
         burning of natural uranium. The dependence of the dimensionless density
	 of uranium-238 nuclei on the spatial coordinate $N_8^* (x)$ for the 
	 moments of time 
	 (a) $t=6~min$; 
     (b) $t = 12~min$;
	 (c) $t = 18~min$; 
	 (d) $t = 24~min$.}
\label{fig:09}
\end{figure}

\begin{figure}[htb]
\centering
\includegraphics[width=4cm]{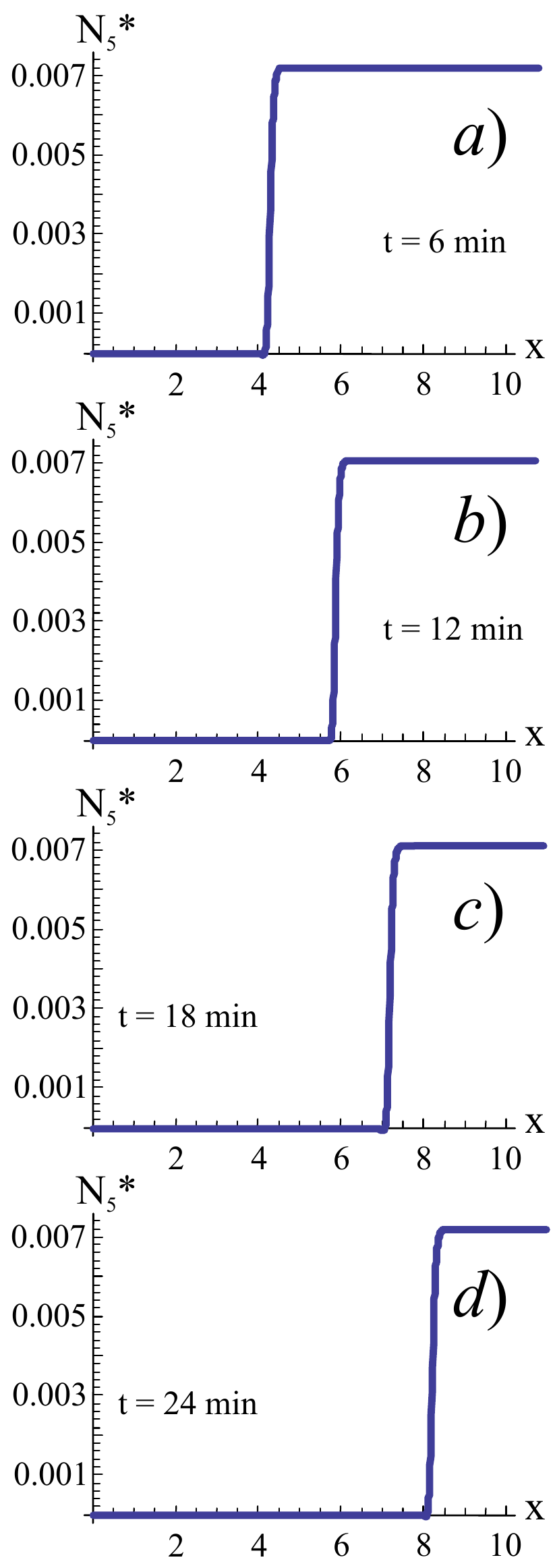}
\caption{Kinetics of the uranium-235 nuclei density under wave neutron-nuclear 
         burning of natural uranium. The dependence of the dimensionless density
	 of uranium-235 nuclei on the spatial coordinate $N_5^* (x)$ for the 
	 moments of time 
	 (a) $t=6~min$; 
	 (b) $t = 12~min$; 
	 (c) $t = 18~min$; 
	 (d) $t = 24~min$.}
\label{fig:10}
\end{figure}

\begin{figure}[htb]
\centering
\includegraphics[width=4cm]{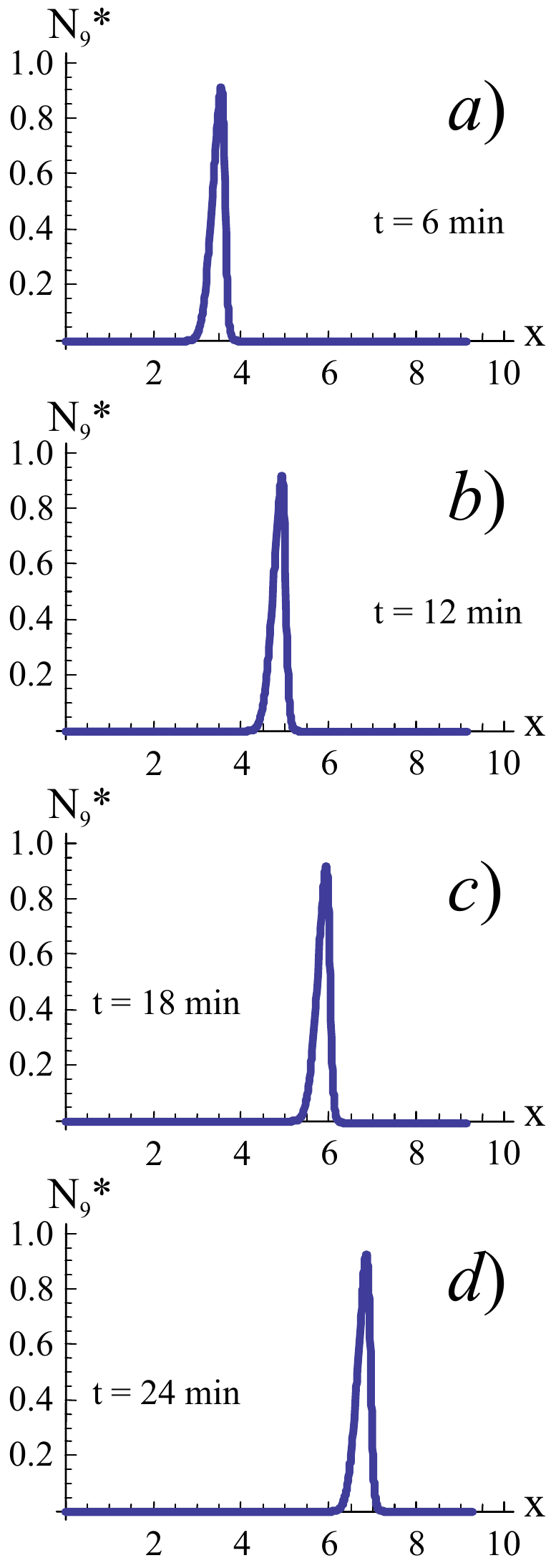}
\caption{Kinetics of the uranium-239 nuclei density under wave neutron-nuclear 
         burning of natural uranium. The dependence of the dimensionless density
	 of uranium-239 nuclei on the spatial coordinate $N_9^* (x)$ for the 
	 moments of time 
	 (a) $t=6~min$; 
	 (b) $t = 12~min$; 
	 (c) $t = 18~min$; 
	 (d) $t = 24~min$.}
\label{fig:11}
\end{figure}

\begin{figure}[htb]
\centering
\includegraphics[width=4cm]{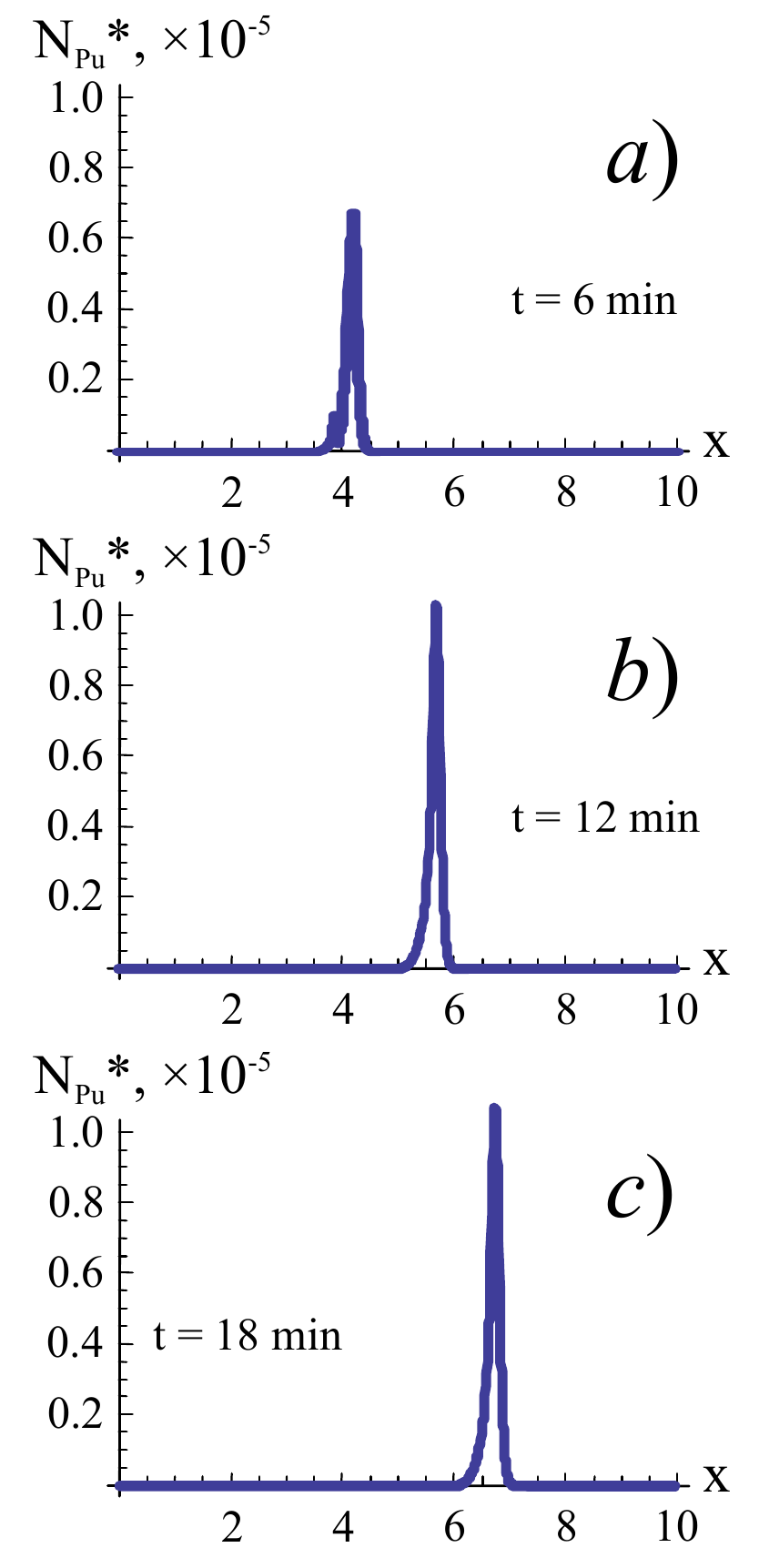}
\caption{Kinetics of the plutonium-239 nuclei density under wave neutron-nuclear
         burning of natural uranium. The dependence of the dimensionless density
	 of plutonium-239 nuclei on the spatial coordinate $N_{Pu}^* (x)$ for 
	 the moments of time 
	 (a) $t=6~min$; 
	 (b) $t = 12~min$; 
     (c) $t = 18~min$.}
\label{fig:12}
\end{figure}

Let us emphasize that according to the results presented in Fig.~\ref{fig:12}, 
the wave of slow neutron-nuclear burning of plutonium-239 has been practically 
formed during the time of modelling, being equal to 18 minutes (the time of the 
wave lighting). The shorter time of the wave lighting in comparison with the 
results of modelling of slow wave burning for fast neutrons (see 
e.g.~\cite{ref03,ref04,ref05,ref10}) is explained by the value of the 
cross-section for the radiation capture of neutrons for uranium-238, set for the
epithermal region amounting to 252.5~barn (see~(\ref{eq50})), being two orders 
higher than the corresponding value for fast neutrons. It should be noted that 
since the cross-section for the radiation capture of neutrons with energy 7~eV 
for uranium-238 has a resonance $\sim 10^4$~barn (see Fig.~\ref{fig:02}), the 
displacement of the maximum of the neutron energy spectrum closer to 7~eV can 
reduce the lighting time or the neutron flux density, created by the external 
(lighting) source by two more orders.

For the second calculation the same (aforesaid) constant coefficients of 
differential equations were set, as for the first calculation, with the 
exception of the effective cross-sections of neutron radiation capture reactions
for fragments and slags. In this calculation the effective cross-sections of the
neutron radiation capture reactions for fragments and slags were increased by 
one order in comparison with the first calculation and had the following values:

\begin{equation}
\sigma_c ^{eff (Pu)} = 10.1 \cdot 10^{-24} ~cm^2; ~ ~
\sigma_c ^{eff (5)} = 10.1 \cdot 10^{-24} ~cm^2: ~ ~
\sigma_c ^{eff} = 10.1 \cdot 10^{-24} ~cm^2: ~ ~
\end{equation}

These values were set according to the data of the base ENDF/B-VII.0 for the 
cross-sections of the neutron radiation capture for products of the fissile 
uranium-plutonium medium for the considered epithermal region of neutron 
energies.

The length of the fissile medium, where the wave of neutron-nuclear burning 
propagates, is equal to $1000~cm$, the full time of modelling is $t = 60 days$, 
the temporal step is $\Delta t = 300~s$, the spatial coordinate step is 
$\Delta x = 0.001~cm$.

The results of the second calculation are presented below in 
figures~\ref{fig:13}-\ref{fig:18}.

\begin{figure}[htb]
\centering
\includegraphics[width=4cm]{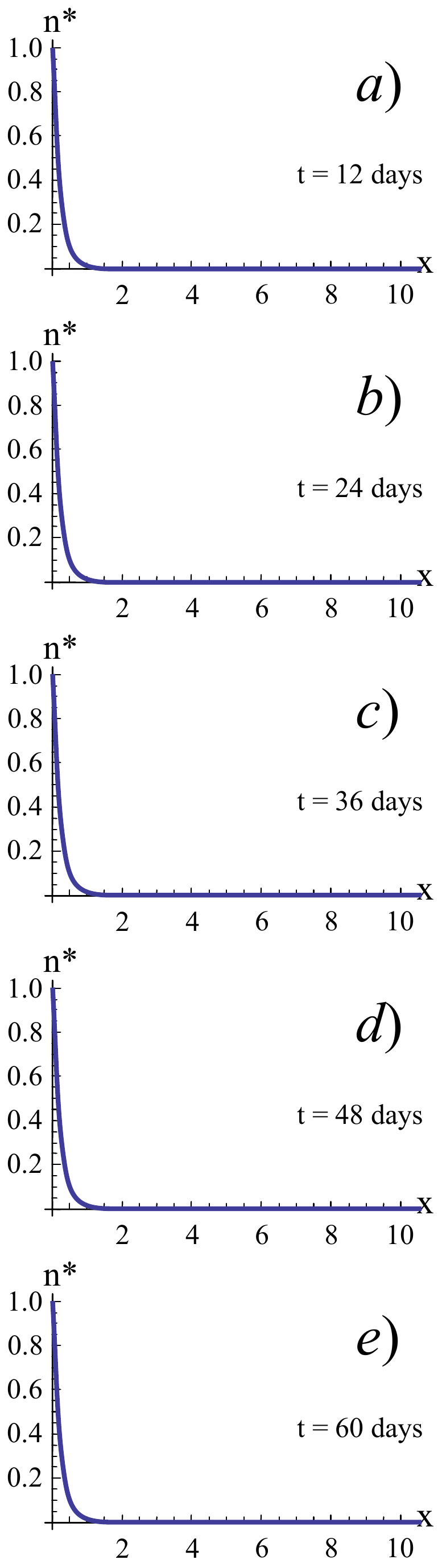}
\caption{Kinetics of neutrons under wave neutron-nuclear burning of natural 
         uranium. The dependence of the dimensionless density of neutrons on 
	 the spatial coordinate $n^* (x)$ for
	 (a) $t = 12~days$; 
	 (b) $t = 24~days$; 
	 (c) $t = 36~days$; 
	 (d) $t = 48~days$;
	 (e) $t = 60~days$).}
\label{fig:13}
\end{figure}

\begin{figure}[htb]
\centering
\includegraphics[width=4cm]{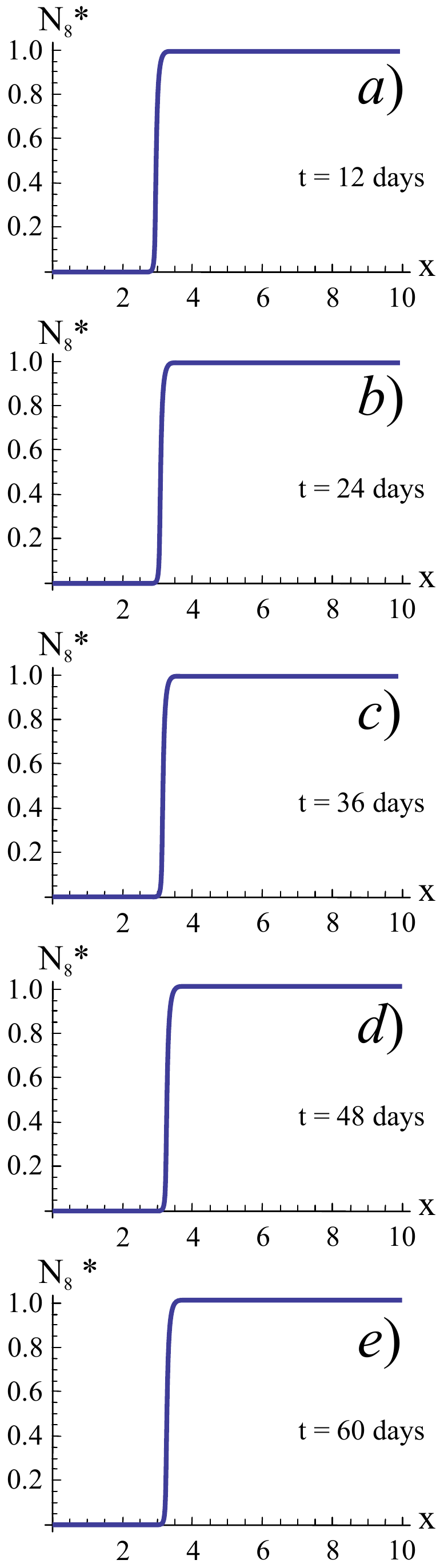}
\caption{Kinetics of the uranium-238 nuclei density under wave neutron-nuclear 
         burning of natural uranium. The dependence of the dimensionless 
	 uranium-238 nuclei density on the spatial coordinate $N_8^* (x)$ for 
	 (a) $t = 12~days$; 
	 (b) $t = 24~days$; 
	 (c) $t = 36~days$; 
	 (d) $t = 48~days$;
	 (e) $t = 60~days$).}
\label{fig:14}
\end{figure}

\begin{figure}[htb]
\centering
\includegraphics[width=4cm]{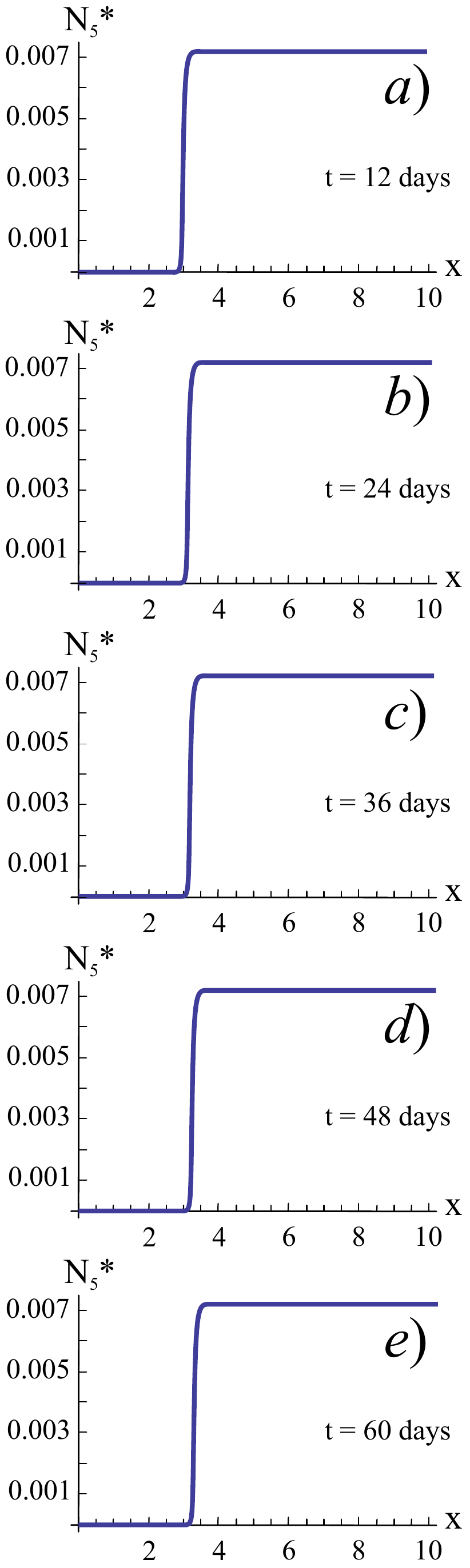}
\caption{Kinetics of the uranium-235 nuclei density under wave neutron-nuclear 
         burning of natural uranium. The dependence of the dimensionless 
	 uranium-235 nuclei density on the spatial coordinate $N_5^* (x)$ for 
	 (a) $t = 12~days$; 
	 (b) $t = 24~days$; 
	 (c) $t = 36~days$; 
	 (d) $t = 48~days$;
	 (e) $t = 60~days$).}
\label{fig:15}
\end{figure}

\begin{figure}[htb]
\centering
\includegraphics[width=4cm]{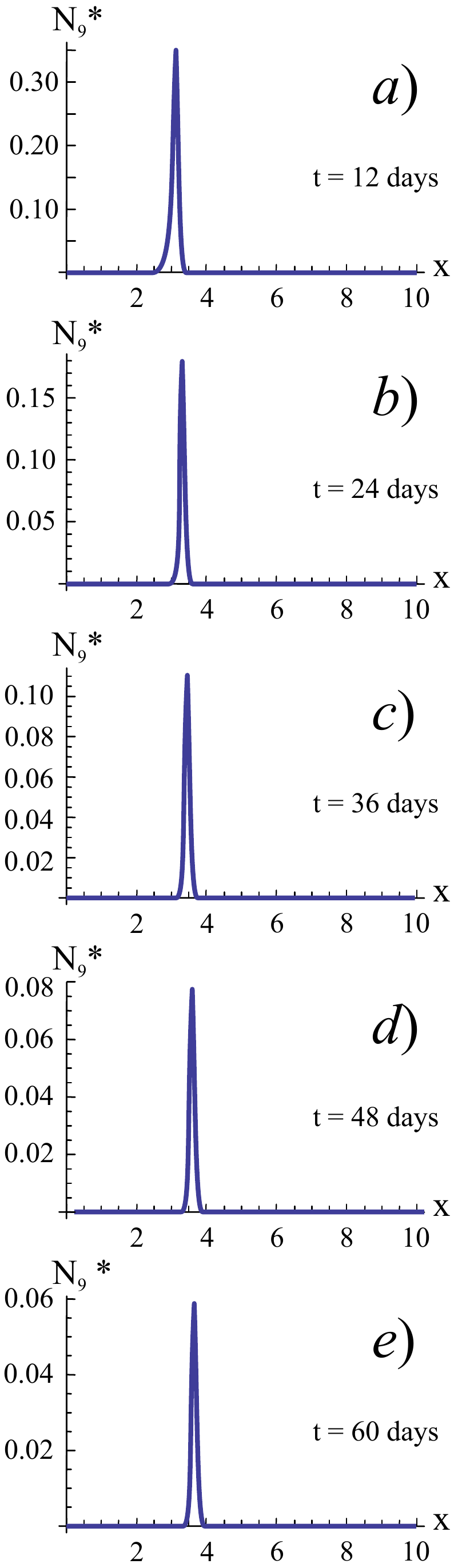}
\caption{Kinetics of the uranium-239 nuclei density under wave neutron-nuclear 
         burning of natural uranium. The dependence of the dimensionless 
	 uranium-239 nuclei density on the spatial coordinate $N_9^* (x)$ for 
	 (a) $t = 12~days$; 
	 (b) $t = 24~days$; 
	 (c) $t = 36~days$; 
	 (d) $t = 48~days$;
	 (e) $t = 60~days$).}
\label{fig:16}
\end{figure}

\begin{figure}[htb]
\centering
\includegraphics[width=4cm]{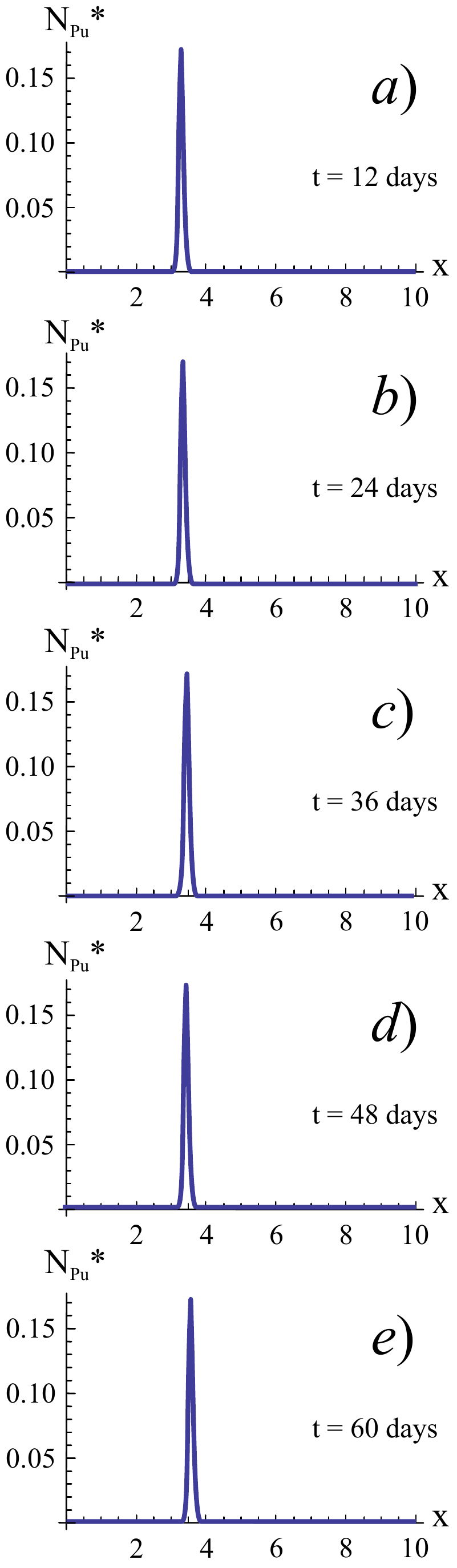}
\caption{Kinetics of the plutonium-239 nuclei density under wave neutron-nuclear
         burning of natural uranium. The dependence of the dimensionless 
	 plutonium-239 nuclei density on the spatial coordinate $N_{Pu}^* (x)$ 
	 for 
	 (a) $t = 12~days$; 
	 (b) $t = 24~days$; 
	 (c) $t = 36~days$; 
	 (d) $t = 48~days$;
	 (e) $t = 60~days$).}
\label{fig:17}
\end{figure}

\begin{figure}[htb]
\centering
\includegraphics[width=4cm]{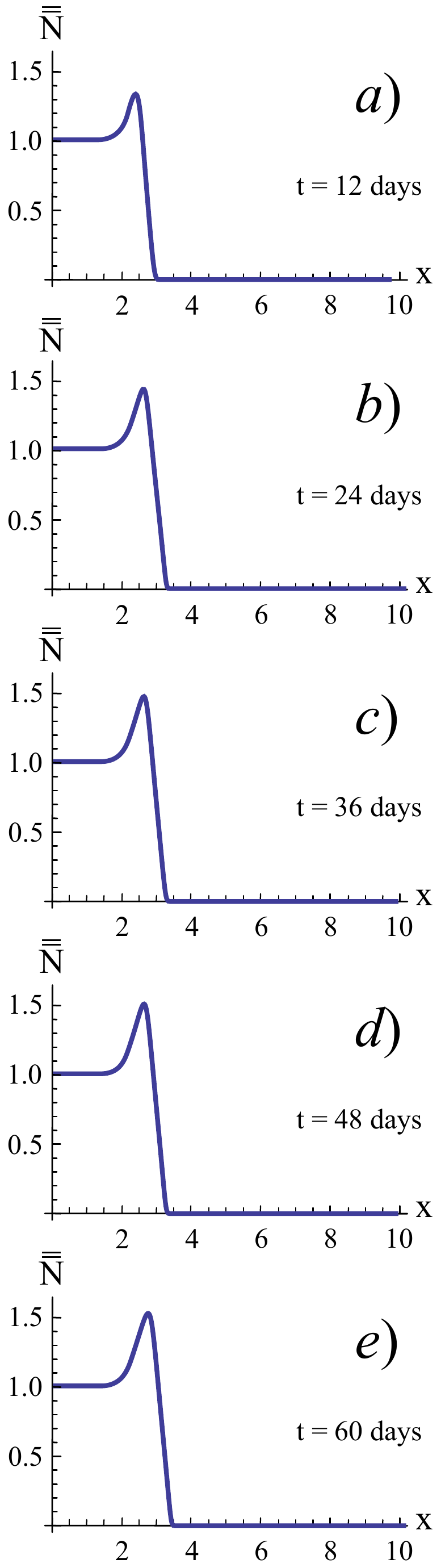}
\caption{Kinetics of the ``slags'' nuclei density under wave neutron-nuclear 
         burning of natural uranium. The dependence of the dimensionless 
	 ``slags'' nuclei density on the spatial coordinate 
	 $\overline{\overline{N}}^* (x)$ for 
	 (a) $t = 12~days$; 
	 (b) $t = 24~days$; 
	 (c) $t = 36~days$; 
	 (d) $t = 48~days$;
	 (e) $t = 60~days$).}
\label{fig:18}
\end{figure}

The presented results of the second numerical modelling of the wave 
neutron-nuclear burning of a natural uranium in the epithermal region of neutron
energies (0.1$\div$7.0~eV) also indicate the realization of such a mode. For 
example, in Fig.~\ref{fig:17} we can see the wave burning of plutonium-239. The 
time of the steady autowave burning establishing for plutonium is 
45$\div$48~days, and the velocity of the steady wave burning of plutonium in 
this case equals $u_numer \approx 1.85 \cdot 10^{-5}~cm/s$ (see 
Fig.~\ref{fig:17}), being two orders smaller than the velocity of the wave 
burning of plutonium for the first numerical calculation (see 
Fig.~\ref{fig:12}). Thus, the increase of the radiation capture effective 
cross-section for slags by one order in comparison with the first calculation 
has led to decrease of the plutonium burning wave velocity by two orders. At the
same time the maximum plutonium concentration in the wave grew up to 15\%.

\section{Estimate of the slow neutron-nuclear burning rate for the thermal
         region of neutron energies}

According to the theory of a soliton-like neutron wave of slow nuclear burning, 
developed on basis of the theory of quantum chaos in~\cite{ref10}, the 
velocities of neutron-nuclear burning must satisfy the Wigner distribution. The 
phase velocity $u$ of the soliton-like neutron wave of nuclear burning is 
determined by the following approximate equality:

\begin{equation}
\Lambda (a_*) = \frac{u \tau_\beta}{2L} \cong \left( \frac{8}{3 \sqrt{\pi}} \right) ^6 a_* ^4 \exp {(-\frac{64}{9 \pi} a_* ^2)}, ~ ~
a_* ^2 = \frac{\pi ^2}{4} \cdot \frac{N_{crit}^{Pu}}{N_{eq}^{Pu} - N_{crit}^{Pu}},
\label{eq52}
\end{equation}

\noindent
where $\Lambda(a_*)$ is a dimensionless invariant, depending on the parameter 
$a_*$; $N_{eq}^{Pu}$ and $N_{crit}^{Pu}$ are the equilibrium and critical 
concentrations of $^{239}Pu$, $L$ is the mean free path of neutrons, 
$\tau_\beta$ is the delay time, connected with birth of an active (fissile) 
isotope and equal to the effective period of the $\beta$-decay of compound 
nuclei in the Feoktistov uranium-plutonium cycle.

For the purpose of checking the velocity of slow neutron-nuclear burning of 
natural uranium in the epithermal region of energies of neutrons 
(1.0$\div$7.0~eV), obtained during numerical modelling, the corresponding 
equilibrium and critical concentrations for plutonium-239 were calculated 
according to the expressions~(\ref{eq28}) and~(\ref{eq29}), and with their help,
using the relationship~(\ref{eq52}), the estimates of the parameter $a_* ^2$ and
the invariant $\Lambda (a_*)$ were made.

In order to calculate the equilibrium and critical concentrations for 
plutonium-239 according to the expressions~(\ref{eq28}) and~(\ref{eq29}), for 
the epithermal region of neutron energies (1.0$\div$7.0~eV) we preliminarily 
calculated the cross-sections of neutron-nuclear reactions averaged over the 
neutron energy spectrum, present in the expressions~(\ref{eq28}) 
and~(\ref{eq29}). The averaging of the neutron-nuclear reactions cross-sections 
in the epithermal region of neutron energies (1.0$\div$7.0~eV) was carried out 
over the neutron spectrum, obtained from the spectrum of WWER neutrons, 
presented in Fig.~\ref{fig:06}, by such its displacement to the epithermal 
region that the maximum of the neutrons spectrum corresponded to their energy 
of 3~eV. The following values were obtained:

\begin{align}
& \bar{\sigma} _c ^{Pu} = 5.43~b, ~ ~
\bar{\sigma} _f ^{Pu} = 26.05~b, ~ ~
\bar{\sigma} _c ^{238} = 251.68~b, ~ ~
\bar{\sigma} _f ^{238} = 0.0001~b, ~ ~ \nonumber \\
& \bar{\sigma} _c ^{235} = 49.41~b, ~ ~
\bar{\sigma} _f ^{235} = 60.19~b, ~ ~
\bar{\sigma} _c ^{239} = 4.68~b, ~ ~
\bar{\sigma} _f ^{239} = 1.64~b; ~ ~ \nonumber \\
& \tilde{N}_{eq}^{Pu} \approx 3.852 \cdot 10^{23} ~1 /cm^3; ~~~ \text{and} ~~~
\tilde{N}_{crit}^{Pu} \approx 4.097 \cdot 10^{22} ~1 /cm^3; \nonumber \\
&a_* = 0.5418 ~~~ \text{and} ~~~ \Lambda(a_*) = 0.5144.
\label{eq53}
\end{align}

The obtained estimates of the parameter $a_*$ and the invariant $\Lambda (a_*)$
for slow neutron-nuclear burning of the natural uranium in the epithermal region
of neutron energies (1.0$\div$7.0~eV) are presented in Fig.~\ref{fig:13}.

At the same time the estimate of the phase velocity of neutron-nuclear burning 
of natural uranium in the epithermal region of neutron energies, obtained by 
means of the numerical modelling results presented in section~\ref{sec05} (see 
e.g. Fig.~\ref{fig:12}) is approximately equal to

\begin{equation}
u_{numer} \approx 1 ~cm/(12 \cdot 60 ~s) \approx 1 \cdot 10^{-3} ~cm/s.
\label{eq54}
\end{equation}

The mean free path for neutrons of the indicated epithermal energy region is equal to

\begin{equation}
L = \frac{1}{\Sigma_a} = \frac{1}{\bar{\sigma}_c ^9 N_8 (t=0)} \approx 
\frac{1}{4.68 \cdot 10^{-24} ~cm^2 \cdot 0.48 \cdot 10^{23} ~cm^{-3}} \approx 
4.45 ~cm.
\label{eq55}
\end{equation}

Here it should be noted that in the previous expression~(\ref{eq54}) for the 
estimation of the averaged free path for neutrons we used the value of the 
radiation capture cross-section for uranium-239 instead of the radiation capture
cross-section for uranium-238, since under the set neutron flux density 
uranium-238 changes to uranium-239 in a very short time as a result of the 
neutron radiative capture reaction.

\begin{figure}[htb]
\centering
\includegraphics[width=10cm]{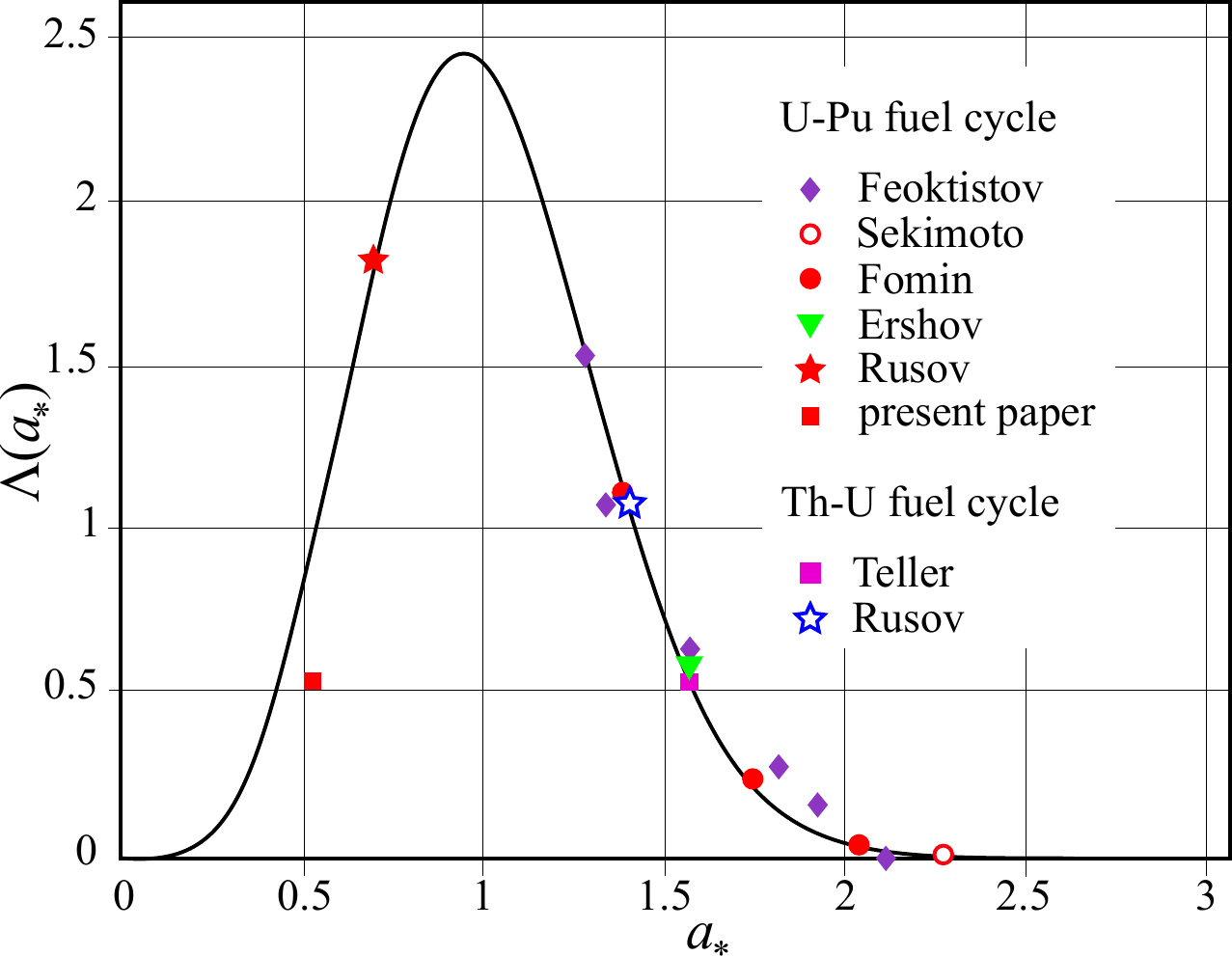}
\caption{The theoretical dependence (the solid line) and the simulation (dots) 
         for phase velocities of neutron-nuclear burning 
	 $\Lambda (a_*) = u \tau_\beta / 2L$ on the parameter $a_*$, presented 
	 in~\cite{ref10} and supplemented with an estimate, obtained in the 
	 present paper for slow wave burning of natural uranium in the 
	 epithermal region of the neutrons energy (1.0$\div$7.0~eV).}
\label{fig:19}
\end{figure}

Thus, on basis of the numerical modeling results we get the following estimate 
of the invariant of nuclear burning:

\begin{equation}
\frac{u_{numer}}{L} \approx \frac{1.00 \cdot 10^{-3} ~cm/s}{4.45~cm} \approx 
2.25 \cdot 10^{-4} ~s^{-1}
\label{eq56}
\end{equation}

The estimate of $u/L$ can be also obtained from the expression~(\ref{eq30}) if 
the values of the invariant $\Lambda$ presented in~(\ref{eq53}) and $\tau_\beta$
are known. Indeed, according to~(\ref{eq52}) we obtain

\begin{equation}
\frac{u}{L} = \frac{2\cdot \Lambda}{ \tau _\beta} \approx 
\frac{2 \cdot 0.51}{2.85 \cdot 10^5 ~s} \approx 3.6 \cdot 10^{-4} ~s^{-1}.
\label{eq57}
\end{equation}

The comparison of the values~(\ref{eq56}) and~(\ref{eq57}) allows to draw a 
conclusion about their good agreement. The difference between the value of the 
invariant~(\ref{eq56}), obtained from the results of numerical modelling, and 
the theoretical estimate~(\ref{eq57}) can be caused by the fact that we used the
one-group diffusion approximation of the neutron transport theory for the 
numerical modelling. Owing to this, we set the values of the neutron-nuclear 
reactions cross-sections~(\ref{eq51}), obtained by averaging over the range of 
epithermal energies (0.1$\div$7.0~eV) for the input data without taking into 
account the form of the neutron energy spectrum. We set the values corresponding
to the energy of 1~eV for the diffusion coefficient and the neutron velocity.

It should be noted that the decrease of the plutonium wave burning velocity in 
the second modelling calculation in comparison with the first one, caused by the
increase of the effective cross-section of the neutron radiation capture for 
fragments of fission and ``slags'' by an order of magnitude, agrees well with 
the expression for the phase velocity of wave burning set by the Wigner 
distribution~(\ref{eq52}) and the expressions~(\ref{eq28}) and~(\ref{eq29}) (or 
the similar ones~(\ref{eq21}) and~(\ref{eq22})) for the equilibrium and critical
plutonium concentrations. The increase of the effective cross-section of neutron
radiation capture for fragments of fission and ``slags'', according to the 
expression~(\ref{eq28}), does not change the estimate for the equilibrium 
plutonium concentration and, according to the expression~(\ref{eq29}), leads to 
the increase of the first term in the numerator of the expression~(\ref{eq29}), 
leading to the increase of the estimate for the critical plutonium 
concentration. The increase of the critical plutonium concentration for the 
constant equilibrium plutonium concentration and under fulfilment of the burning
condition $\tilde{N}_{eq}^{^{239}_{94}Pu} > \tilde{N}_{crit}^{^{239}_{94}Pu}$, 
according to the expression~(\ref{eq52}) for the parameter $a_*$, leads to its 
increase, since the numerator grows up while the denominator goes down, and such
parameter increase causes the decrease of the phase velocity of plutonium wave 
burning also according to the expression~(\ref{eq52}).

\section*{Conclusions}

A general criterion of the wave modes of neutron-nuclear burning realization for
both Akhiezer and Feoktistov waves is formulated for the first time.

The investigation of the wave burning criterion fulfilment for a fissile medium 
originally consisting of uranium-238 in a wide range of neutron energies is 
conducted for the first time. The possibility of the wave nuclear burning not 
only in the region of fast neutrons, but also for cold, epithermal and resonance
ones is also discovered for the first time.

The results of the investigation of the Feoktistov criterion fulfilment for a 
fissile medium, originally consisting of uranium-238 dioxide with enrichments 
4.38\%, 2.00\%, 1.00\%, 0.71\% and 0.50\% with respect to uranium-235, in the 
region of neutron energies 0.015$\div$10.00~eV are presented for the first time.
These results indicate a possibility of the ultraslow wave neutron-nuclear 
burning of the uranium-plutonium media, originally (before the wave mode 
initiation with some external neutron source) having enrichments with respect to
uranium-235, corresponding to the subcritical state, in the regions of cold, 
thermal, epithermal and resonance neutrons.

In order to confirm the validity of the conclusions based on the analysis of the
slow wave neutron-nuclear burning criterion fulfilment depending on the neutron 
energy, the numerical modelling of the ultraslow wave neutron-nuclear burning in
the natural uranium within the epithermal region of neutron energies 
(0.1$\div$7.0~eV) was conducted for the first time. The presented results of the
numerical modelling of such conditions indicate the realization of such a mode.

A conclusion about the possibility of creating the thermal-epithermal wave 
nuclear reactor in which a natural uranium in its various forms can be used as 
fuel, substantiated by the calculations results, is made for the first time. We 
also make a conclusion that light-water thermal reactors with natural uranium 
fuel are appropriate for slow wave burning of plutonium-239. In a heavy-water 
natural uranium reactor there is no region, where the critical plutonium 
concentration is positive and, consequently, the mode of slow wave 
neutron-nuclear burning (the Feoktistov wave) cannot be realized. The same is 
true for the natural uranium reactors with a gas coolant. In these reactors slow
wave nuclear burning is possible for fuel with so small enrichment with respect 
to uranium-235, that in the thermal energy region the plutonium-239 criticality 
region exists. For example, engineering uranium or already spent nuclear fuel 
with appropriate burn-up will do.

Thus we make the conclusion about a possibility of creation of wave nuclear 
reactors on cold, thermal, epithermal and resonance neutrons, and not only on 
fast neutrons for the first time, which is substantiated by the corresponding 
calculations. It is extremely important for a number of reasons, the major ones 
of which are the following. First, the problem of materials radiation 
resistance, being topical for fast wave reactors~\cite{ref20}, is resolved 
automatically, since the neutron energy is smaller by approximately six orders 
(from $\sim$1~MeV to $\sim$1~eV) and, consequently, the integral fluence on the 
material of the fuel element walls also much smaller. However, we would like to 
draw attention to the fact that, as it follows from the results presented in 
Fig.~\ref{fig:05}, the wave burning criterion holds true not only for fast 
reactors with the hard spectrum (the average neutron energy $\sim$1~MeV), but 
also for fast reactors with the softened neutron spectrum (the average neutron 
energy less then $\sim$100~keV, and a ``tail'' of the neutron spectrum with the 
energy less than 10~keV is rather large, see e.g.~\cite{ref17}). And the 
solution of the problem of the materials radiation resistance, essential for 
fast wave reactors with the hard neutron spectrum, requires creation of new 
structural materials for the fuel element walls, withstanding the radiation load
of 500~dpa (displacement per atom), whereas the materials operating nowadays 
withstand $\sim$100~dpa (see~\cite{ref20}), i.e. the increase of radiation 
resistance in five times is necessary. Therefore, perhaps, it will be possible 
to solve the existing problem of radiation resistance of materials for fast wave
reactors with the hard neutron spectrum by creating the fast wave reactors with 
the softened neutron spectrum, since the transition to such reactors reduces 
the radiation load by an order. Let us also note that the prohibition on usage 
of steel as a constructional material of the active zone and necessity of use 
the expensive aluminium and zirconium alloys instead, which takes place for 
usual thermal reactors operating in the aforesaid ranges of neutron energies, 
do not exist for the wave reactors, since there is no need to preserve the 
neutrons for maintaining the chain reaction in wave reactors. Consequently, a 
possibility to exclude the danger of zirconium-steam reaction with the 
subsequent explosion of the hydrogen mixture~\cite{ref15,ref21,ref22} appears. 
Second, the wave reactors on resonance neutrons can act as the transmutators of 
the nuclides most dangerous for biosphere, being fragments of the fuel nuclides 
fissions (the so-called biocompatibility of the nuclear reactor of new 
generation). Third, the realization of the reactors on cold neutrons simplifies 
their radiation protection seriously because of small penetrability of the 
neutrons with these energies, which in combination with internal safety of the 
wave reactors can ensure their wide spread adoption. Fourth, a possibility of 
the wave reactor realization in the epithermal neutron region (the maximum of 
neutron distribution of the Maxwellian type in the region from 3~eV to 7~eV) 
attracts profound interest, since the requirements for the flux density of the 
external neutron source, ensuring the lighting of the nuclear burning, decrease 
sharply in this case (approximately by three-four orders). This is also true for
the lighting times, because of the presence of the radiation capture 
cross-section maximum for uranium-238 of the order of 10000~barn for the neutron
energy 7~eV. A source with natural radioactivity of neutrons with the energy 
7~eV could be such a source, as well as the accelerator of charged particles 
(e.g. protons or electrons), which create mainly neutrons with the energy 7~eV 
by interaction with the target nuclei as a result of the nuclear reaction.

Let us note that wave nuclear burning in the epithermal region of neutron 
energies is of serious interest also for investigation of the possible burning 
modes of the wave georeactor~\cite{ref05}, since the epithermal region of 
neutron energies with the distribution maximum for neutrons of Maxwellian type 
in the region $\sim$1~eV can be considered as a region of thermalized neutrons 
in the fissile uranium-plutonium medium, being in a state with the temperature 
$\sim$5000~K, which corresponds to the temperature of nuclear burning of the 
georeactor at the interface of the solid and liquid Earth cores.

\section*{Acknowledgements}

M.V. Eingorn acknowledges support by NSF CREST award HRD-0833184 and NASA grant 
NNX09AV07A and thanks Prof. Branislav Vlahovic (NCCU) for the given 
computational resources without which the numerical calculation would be 
impossible on the same level of accuracy.


\begin{thebibliography}{99}

\bibitem{ref01}
    Feoktistov, L.P. Neutron-fission wave, Dokl. Akad. Nauk, Vol. 309, 1989.

\bibitem{ref02}
    Akhiezer A.I., Belozorov D.P., Rofe-Beketov F.S., Davydov L.N., Spolnik Z.A.
    On the theory of propagation of chain nuclear reaction in diffusion approximation, Yad. Fiz., Vol.62, 1999, 1567-1575.

\bibitem{ref03}
    V.D. Rusov, V.A. Tarasov, D.A. Litvinov, \textit{Reactor antineutrino 
    physics (in Russian)}, Moscow, URSS, 2008.

\bibitem{ref04}
    Rusov V.D., Tarasov V.A., Vaschenko V.N. Traveling wave nuclear reactor -- Kyiv: Publishing group "A.C.C.", 2013. --  156 p.

\bibitem{ref05}
    Rusov V.D., Pavlovich V.N., Vashenko V.N., Tarasov V.A. , et al. Geoantineutrino spectrum and slow nuclear burning on the boundary of the liquid and solid phases of the Earth's core // Journal of Geophysical Research. 2007. Vol. 112, B09203, doi: 10.1029/2005JB004212. P. 1-16.

\bibitem{ref06}
    A.I.Akhiezer and I.Ya.Pomeranchuk. Some Problems of Nuclear Theory. Moscow: Gostechizdat, 1950 (in Russian)
    

\bibitem{ref07}
    Glasstone, S. (1952). The elements of nuclear reactor theory. New York: Van Nostrand.

\bibitem{ref08}
    S. M. Feinberg, S. B. Shikhov, and V. B. Troyanskii,Theory of Nuclear Reactors, Volume 1, Elementary Theory of Reactors, A College Textbook for Colleges [in Russian], Atomizdat, Moscow (1978).

\bibitem{ref09}
    	 S. V. Shirokov. “Nuclear Reactor Physics (in Russian),” Naukova Dumka, Kiev, 1992.

\bibitem{ref10}
    Rusov V.D., Linnik E.P., Tarasov V.A., et al. Traveling Wave Reactor and Condition of Existence of Nuclear Burning Soliton-like Wave in Neutron-Multiplying Media //  Energies (Special Issue “Advances in Nuclear Energy”), 4 (2011), p. 1337-1361.

\bibitem{ref11}
   L.D. Landau, E.M. Lifshitz (1987). Fluid Mechanics. Vol. 6 (2nd ed.). Butterworth-Heinemann. ISBN 978-0-08-033933-7.

\bibitem{ref12}
Zeldovich, Ia. B. Physics of Shock Waves and High-Temperature Hydrodynamic Phenomena, Vol. 1 \& 2. New York: Academic Press, 1966, 1967.

\bibitem{ref13}
    Rusov, V.D. and Tarasov, V.A. and Chernegenko, S.A., Blow-up modes in uranium-plutonium fissile medium in technical nuclear reactors and georeactor (in Russian), Problems of Atomic Science and Technology, Vol.72, Series: Physics of Radiation Effect and Radiation Materials Science, Vol.97 , 2011, pp.123-131.

\bibitem{ref14}
    V.D. Rusov, V.A. Tarasov, S.I. Kosenko, S.A. Chernegenko, The resonance absorption probability function for neutron and multiplicative integral, Problems of Atomic Science and Technology, Vol.78, 2012, pp.112-121. arXiv:1208.1019v1 [nucl-th].

\bibitem{ref15}
    Rusov V.D., Tarasov V.A., Vaschenko V.M., Linnik E.P., Zelentsova T.N., Beglaryan M.E., Chernegenko S.A., Kosenko S.I., Molchinikolov P.A., Smolyar V.P., Grechan E.V. Fukushima plutonium effect and blow-up regimes in neutron-multiplying media. // World Journal of Nuclear Science and Technology, 2013, No.3, p. 9-18; arXiv:1209.0648v1 [nucl-th].

\bibitem{ref16}
     Yu. M. Shirokov, N. P. Yudin, Nuclear Physics, Imported Pubn (Jun 1983).

\bibitem{ref17}
G.G. Bartolomey, G.A Bat', V.D. Baibakov, and M.S. Altukhov. Basic theory and methods of nuclear power installations calculation. Energoatomizdat, Moscow, 1989, 512p. (in Russian)

\bibitem{ref18}
    V. I. Vladimirov Physics nuclear reactors Practical problems in their operation, Energoatomizdat, Moscow, 1986, 304p. (in Russian)

\bibitem{ref19}
    A.I. Akhiezer, I.Ya. Pomeranchuk. Introduction into the theory of neutron multiplicating systems (reactors), IzdAT, Moscow, 2002.


\bibitem{ref20}
    Rusov V. D., Tarasov V. A., Sharf I. V.,  Vaschenko V. M., Linnik E. P., Zelentsova T. N., Beglaryan M. E., Chernegenko S. A., Kosenko S. I., Molchinikolov P. A., Smolyar V. P.  and Grechan E. V. On some essential peculiarities of the traveling wave reactor operation. // arXiv:1207.3695 [nucl-th].

\bibitem{ref21}
    Rusov V.D., Tarasov V.A., Chernezhenko S.A., Kakaev A.A., Grechan E.V., 
Kosenko S.I., Pantak O.I. \textit{The temperature dependences distinction of 
thermal source densities of MOX-fuel and dioxide-fuel and related with it the 
features of the AES ``Fukusima-1'' third block accident}, \textbf{Proc. Int. 
Conf. Current Problems in Nuclear Physics and Atomic Energy} (NPAE-Kyiv2012). 
2012, September 10–14, Kyiv, Ukraine. P.479-483.

\bibitem{ref22}
    Rusov V.D., Tarasov V.A., Chernezhenko S.A., Kakaev A.A., Grechan E.V., 
\textit{A criterion of the wave nuclear burning in the U-Pu fissile medium
and a thermal neutron spectrum of the WWER}, \textbf{Proceedings of the 3$^{rd}$
International Scientific and Technical Conference ``Towards a higher safety and 
efficiency of atomic energy''}, Odessa, Ukraine, September 24-28, 2012, 
pp.189-202. (in Russian).


\end{thebibliography}
\end{document}